\documentclass[aps,prl,twocolumn,amsmath,amssymb,superscriptaddress]{revtex4-1}

\usepackage{graphics}
\usepackage{dcolumn}
\usepackage{bm}
\usepackage[caption=false]{subfig} 
\usepackage{color}
\usepackage{tikz}
\usepackage{subfig}
\usepackage[titles,subfigure]{tocloft}
\usepackage{soul}

\usepackage[titletoc,toc,title]{appendix}

\DeclareMathAlphabet\mathbfcal{OMS}{cmsy}{b}{n}

\begin{document}

\title{A Passage to Topological Matter: Colloquium}

\author{Kwon Park}
\email[Electronic address:$~~$]{kpark@kias.re.kr}
\affiliation{School of Physics, Korea Institute for Advanced Study, Seoul 02455, Korea}

\date{\today}

\begin{abstract}
Topological matter has become one of the most important subjects in contemporary condensed matter physics.
Here, I would like to provide a pedagogical review explaining some of the main ideas, which were pivotal in establishing topological matter as such an important subject.
Specifically, I explain how the integer quantum Hall state played the role as a prototype for topological insulator, eventually leading to the concept of topological matter in general. 
The topological nature of the integer quantum Hall state is best represented by the Thouless-Kohmoto-Nightingale-den Nijs, or so-called TKNN formula, which connects between the Berry phase and the Hall conductivity. 
The topological non-triviality of topological insulator stems from the existence of a Dirac monopole in an appropriate, but often hidden Hamiltonian parameter space.
Interestingly, having the identical Dirac monopole structure, the Hamiltonian describing the Rabi oscillation bears the essence of topological insulator.
The concept of topological matter has expanded to include topological semimetals such as Weyl and Dirac semimetals. 
A final frontier in the research of topological matter is the interaction-induced topological phases of matter, namely, the fractional Chern and topological insulators. 
The existence of the fractional Chern and topological insulators has been proposed theoretically by drawing an analogy from the fractional quantum Hall states.  
The gist of this proposal is explained along with some of its issues.
I conclude this review by discussing some of the future directions in the research of topological matter.
\end{abstract}

\maketitle


There are various reasons why topological matter has become one of the most important subjects in contemporary condensed matter physics.
While the specific ordering of importance can vary depending on the authors, one of the key reasons should be the fact that topological matter owes its existence to the Berry phase~\cite{Berry1984}.
Prior to the establishment of topological matter, it was believed that the Berry phase could manifest itself only through a form of interference. 
The notion that the Berry phase can be used as a new order parameter defining topological matter has received a universal recognition as one of the most important breakthroughs in condensed matter physics.

In this review, I would like to explain how this single idea of using the Berry phase as a topological order parameter has generated a remarkable chain of discoveries and new ideas, establishing topological matter as one of the most important subjects.

{\bf Berry phase as a topological order parameter}

While the usual order parameter is related with a spontaneous breaking of the symmetry in the Hamiltonian, the {\it topological order parameter} has to do with the topological structure of the manifold formed by the Hamiltonian parameter.

To give an intuitive example illustrating what this means physically, let us imagine a Gedanken experiment of the Aharonov-Bohm effect with an infinitesimally thin solenoid. 
As well known, electrons traveling along a certain closed path would acquire the Aharonov-Bohm phase depending on whether the path encloses the solenoid or not.
This already indicates some form of topology.

Generally speaking, however, the total flux penetrating through the solenoid is not quantized if the solenoid is indeed completely shielded from the electron paths.
Therefore, although the Aharonov-Bohm phase acquired by each electron path is topological in the sense that it depends on whether the path encloses the solenoid or not, the total flux itself is not topologically quantized. 
Fortunately, there is a class of superconductor known as the type-II superconductor, where an externally applied magnetic field can penetrate the superconductor as a lattice of narrow bundles.
This magnetic bundle is called the magnetic vortex since it induces a swirling supercurrent of Cooper pairs nearby.  
Now, the total flux of the magnetic vortex is quantized due to the condition that the Cooper-pair wave function should be single valued circling around each vortex. 
This means that any closed electron path can be topologically classified in terms of how many magnetic vortices it encloses.   
In other words, the number of vortex quanta inside a closed path can be used as a topological order parameter of the path.

The topological insulator~\cite{Hasan2010,Qi2011_Review} can be regarded as a generalization of this idea to higher dimensions. 
In a broad sense, the term ``topological insulator'' includes both the anomalous quantum Hall state, also known as the Chern insulator, and its appropriately extended version with time-reversal symmetry.
In 2D, the time-reversal invariant topological insulator is simply two independent copies of the Chern insulator, preserving the time-reversal symmetry as a whole.

To begin, let us discuss the 2D Chern insulator, which is rather a direct generalization of the 1D idea sketched above for the following reasons. 
First, a closed path in the 1D example corresponds to a compact 2D manifold formed by momenta, i.e., the 2D Brillouin zone, which can be mapped onto the surface of a Bloch sphere in the Hamiltonian parameter space.
Second, a magnetic vortex corresponds to a Dirac monopole generating a hedgehog-like configuration of the effective magnetic field in the Hamiltonian parameter space.  
It is important to note that, similar to the 1D example, the strength of the Dirac monopole is also quantized due to the single-valuedness of the electron wave function. 

{\bf Integer quantum Hall state: Prototype of the Chern insulator}

Historically, the concept of the 2D Chern insulator has been hidden all along in the integer quantum Hall state (IQHS), waiting to be discovered. 
The quantized Hall conductance of the IQHS was so precise that many researchers believed that there must be a fundamental reason for this. 
There have been several different approaches including the Laughlin's gauge argument~\cite{Laughlin1981}, the Thouless-Kohmoto-Nightingale-den Nijs, or so-called TKNN formula~\cite{TKNN1982}, the Landauer-type argument using the edge state transport~\cite{Halperin1982, Streda1987}, and so on.
Different approaches are useful in their own specific purposes.
In this review, we focus on the first two approaches, namely, the Laughlin's gauge argument and the TKNN formula.

{\bf Laughlin's gauge argument.}
The Laughlin's gauge argument is both elegant and powerful since it is based on one of the most fundamental principles in physics, i.e., the gauge invariance principle.
To explain the quantized Hall conductance of the IQHSs, Laughlin performed a Gedanken experiment imagining a very large cylinder of the 2D electron gas (2DEG) system, which can be obtained by connecting two ends of the 2DEG system so that the periodic boundary condition can be applied along the circumference direction of the cylinder.

\begin{figure}
\includegraphics[width=0.7\columnwidth,angle=0]{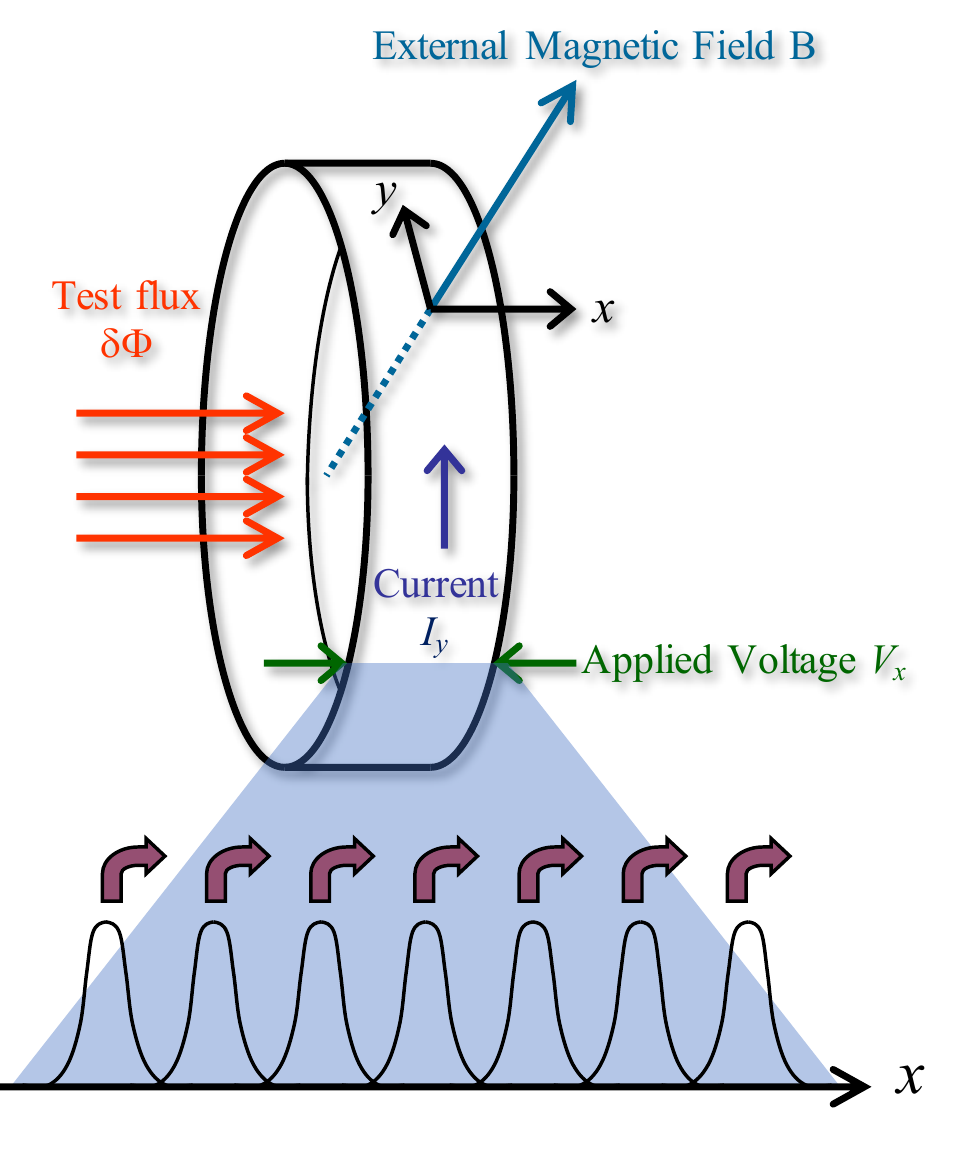}
\caption{Laughlin's gauge argument for the quantized Hall resistance.
In the Landau gauge, the Landau-level eigenstates are the Gaussian-localized wave packets along the $x$ direction, whose center positions are dependent on the momentum in the $y$ direction.
Inserting a test flux $\delta \Phi$ induces a kick in the momentum in the $y$ direction and thus a shift in the center position of the wave packets.
This generates a pumping of the charge along the $x$ direction and a flow of the current along the $y$ direction, eventually explaining the quantized Hall resistance. 
See the main text for details.
}
\label{fig:Laughlin_gauge_argument}  
\end{figure}   

To understand the Laughlin's gauge argument in concrete details, let us consider the Hamiltonian in the Landau gauge ${\bf A}=(0,Bx,0)$:
\begin{align}
H &= \frac{1}{2m} \left[ p_x^2 +\left( p_y-\frac{eB}{c}x\right)^2 \right] 
\nonumber \\
&= \frac{1}{2m} p_x^2 +\frac{1}{2}m\omega_c^2 (x-k_y l_B^2)^2 ,
\label{eq:H_Landau_gauge}
\end{align}
where the $x$ direction is across the cylinder height, and the $y$ direction is along the circumference of the cylinder.  
The external magnetic field ${\bf B}$ is assumed to be applied in such a way that it penetrates the surface of the cylinder perpendicularly. 
Here, $\omega_c=eB/mc$ is the cyclotron frequency, and $l_B=\sqrt{\hbar c /eB}$ is the magnetic length.
Also, the second line is obtained by using the separation of variables, i.e., the wave function is chosen to be the momentum eigenstate along the $y$ direction, $\psi(x,y)=\phi(x) e^{ik_y y}$ with $k_y=2\pi m_y/L_y$.
The Hamiltonian in Eq.~\eqref{eq:H_Landau_gauge} is nothing but the Hamiltonian of the 1D harmonic oscillator, which can be solved readily by the following eigenstates:
\begin{align}
\psi_{n,k_y}(x,y) = {\cal C}_n H_n(x/l_B-k_y l_B) e^{-\frac{(x-k_y l_B^2)^2}{2 l_B^2}} e^{i k_y y} ,
\label{eq:LL_eigenstates}
\end{align}
where $H_n$ is the Hermite polynomial and ${\cal C}_n$ is the normalization constant. 
It is important to note that the center of the wave packet along the $x$ direction, described by the Hermite polynomial times the Gaussian function, is dependent on the $y$-direction momentum, $k_y$.

Returning back to the Laughlin's gauge argument, let us imagine that we now insert a small test flux, $\delta \Phi$, through the cylinder along its axis.  
The insertion of such a test flux induces a kick in $k_y$ due to the Aharanov-Bohm effect, i.e., $k_y \rightarrow k_y+\delta k_y = 2\pi m_y/L_y +\frac{e}{\hbar c} \delta \Phi/L_y$.
This kick in $k_y$ generates in turn a shift in the center of the wave packet to the $x$ direction.
If $\delta \Phi$ becomes one flux quantum $\phi_0=2\pi\hbar c/e$, $\delta k_y=2\pi/L_y$, meaning that the wave packets are shifted exactly by one unit.
Actually, this can be regarded as a form of the gauge transformation since this shift can be absorbed into $m_y$ by defining a new momentum index $m_y^\prime=m_y+1$, which just relabels the wave packets without changing the physical nature of the state.
See Fig.~\ref{fig:Laughlin_gauge_argument} for a schematic diagram.

Now, any IQHS with integer-filled Landau levels should be invariant with respect to this gauge transformation of adding a test flux equal to one flux quantum. 
Interestingly, while this gauge transformation does not change any physical nature of the state itself, something has actually changed. 
What has changed is that charges have been pumped from the left to right end of the cylinder. 
To be precise, when the number of filled Landau levels is $n$, $n$ charges are pumped.
In this situation, if the voltage $V_x$ is applied across the height of the cylinder, then there is an energy increase in the amount of $neV_x$.

Meanwhile, this energy increase due to the flux change should be related to the current, $I_y$, flowing along the circumference of the cylinder.
To understand this, note that the cylinder has a magnetic moment equal to $\mu=I_y A/c$, where $A$ is the area of the cross section of the cylinder. 
Then, the energy increase due to the flux change is given by $\delta E=\mu \delta \Phi/A=I_y \delta \Phi/c$, meaning that $I_y=c \delta E/\delta \Phi$.
Now, let us remind ourselves that, for the IQHS with $n$-filled Landau levels, $\delta E=neV_x$ for $\delta \Phi=\phi_0=2\pi \hbar c/e=hc/e$.
Consequently, the Hall resistance is given as
\begin{align} 
R_{\rm H}=\frac{V_x}{I_y}=\frac{h}{n e^2} ,
\end{align}
which is exactly the expected result for the quantized Hall resistance of the IQHS at $\nu=n$.
Note that the Hall conductance is simply the inverse of the Hall resistance for the IQHSs, where the longitudinal resistance is zero.

At this point, it is very important to note that the above argument should hold even in the presence of moderately strong disorder. 
Moderately strong disorder would disturb the ideal positions of the Landau-level eigenstates.
Despite this disturbance, however, the gauge transformation must move each wave packet one by one, making the whole state return to itself.  
In conclusion, the Hall resistance should be quantized as before.

Actually, there is a tricky issue regarding the role of disorder.
The gauge transformation shifts only the center positions of the so-called extended states, which are extended across the entire circumference of the cylinder. 
On the other hand, the localized states, which are localized near impurities, do not contribute to the Hall resistance since the test flux would have negligible effects on them.
This means that, even though the number of electrons changes continuously, only those electrons forming the extended states can contribute to the Hall resistance.  
Put in another way, any electron spilled over the top-most filled Landau level is trapped in the localized states, making the Hall resistance fixed for a range of the magnetic field.
Therefore, ironically, the Hall resistance can become quantized because of (moderately strong) disorder.
Without disorder, there would be no localized state at all.
In such an idealized situation, the Hall resistance would follow a straight line as a function of filling factor since every electron can contribute.

Of course, this scenario breaks down if disorder becomes too strong.
In this situation, the gap between different Landau levels closes, and therefore there would be no protection against the excitations to higher Landau levels, destroying the quantized Hall resistance.

{\bf TKNN formula.}
Perhaps, the TKNN approach has been the most influential one among various approaches explaining the quantized Hall resistance since it makes a direct connection between the Berry phase and the Hall conductance. 
Despite the conceptual breakthrough, however, the TKNN approach may be the most traditional approach since it uses the linear response theory.

The linear response theory is also known as the fluctuation-dissipation theorem, which is one of the most fundamental principles in condensed matter physics.
In some sense, it is the fluctuation-dissipation theorem that actually makes the comparison between theory and experiment possible.  
The gist of the fluctuation-dissipation theorem is that the dissipation, or response induced by an external perturbation is proportional to the fluctuation, or correlation of the system that is already present in equilibrium, i.e., before the application of the external perturbation. 
The reason why this makes the comparison between theory and experiment possible is that theory can usually compute only the equilibrium properties of the system, while experiment must apply some kind of probe to the system, i.e., an external perturbation.
A well-known example of the fluctuation-dissipation theorem is that the density-density correlation function in equilibrium is proportional to the dielectric function, which is simply the linear response to an external electric field.
See Table~\ref{tab:fluc-diss} for other examples.

\begin{table*}
\centering
\begin{ruledtabular}
\begin{tabular}{llll}
& Fluctuation or correlation & Dissipation or response & \\
\hline
& Density-density & Dielectric function & \\
& Spin-spin & Spin susceptibility & \\
& Current-current along the same direction & Longitudinal conductivity & \\
& Current-current between orthogonal directions & Hall conductivity & \\
\end{tabular}
\end{ruledtabular}
\caption{Examples of the fluctuation-dissipation theorem.
\label{tab:fluc-diss}}
\end{table*}

As shown in Table~\ref{tab:fluc-diss}, it is necessary to consider the current-current correlation between orthogonal directions to compute the Hall conductivity $\sigma_{xy}$.
Mathematically, 
\begin{align}
\sigma_{xy} \propto \int^{\infty}_{0} dt \langle J_x(t) J_y(0) \rangle ,
\label{eq:sigma_xy}
\end{align}
where $J_x$ and $J_y$ are the currents along the $x$ and $y$ directions, respectively.
It turns out that it is possible to perform the integral in the right-hand side of Eq.~\eqref{eq:sigma_xy} exactly for 2D noninteracting systems, as done by TKNN.
After performing some algebra, one can rewrite Eq.~\eqref{eq:sigma_xy} as follows:
\begin{align}
\sigma_{xy} \propto \int d^2 {\bf k} \sum_{\epsilon_{\mu{\bf k}}<\epsilon_F} \langle \nabla_{\bf k} u_{\mu{\bf k}} | \times | \nabla_{\bf k} u_{\mu{\bf k}} \rangle \cdot \hat{z} ,
\end{align}
where, being the periodic part of the Bloch wave function, $u_{\mu{\bf k}}({\bf r})$ is defined as the eigenstate of the following modified Schr\"{o}dinger equation with the eigenvalue $\epsilon_{\mu{\bf k}}$:
\begin{align}
H_{\rm mod} =
\frac{1}{2m}\left(-i\hbar\nabla+\hbar{\bf k}-\frac{e}{c}{\bf A}({\bf r})\right)^2+U({\bf r}) ,
\label{eq:H_mod}
\end{align}
where $U({\bf r})$ is the potential energy imposing the periodic structure of the lattice.
Above, $\hat{z}$ denotes the unit vector along the $z$ direction, i.e., the perpendicular direction to the 2DEG system.
It is important to note that the summation is taken over all the energy bands, whose energy is below the Fermi energy. 
After taking into account the proportionality constant precisely, the Hall conductivity can be written as follows:
\begin{align}
\sigma_{xy}= \frac{e^2}{h} \sum_{\epsilon_{\mu{\bf k}}<\epsilon_F} {\cal C}_\mu ,
\label{eq:sigma_xy2}
\end{align}
where ${\cal C}_\mu$, called the Chern number of the $\mu$-th energy band, is defined by
\begin{align}
{\cal C}_\mu =\frac{i}{2\pi} \int d^2 {\bf k} \langle \nabla_{\bf k} u_{\mu{\bf k}} | \times | \nabla_{\bf k} u_{\mu{\bf k}} \rangle \cdot \hat{z} .
\label{eq:Chern_number}
\end{align}

Equation~\eqref{eq:Chern_number} shows that the Chern number of a given energy band is the total Berry flux piercing through the entire Brillouin zone for that energy band.
To understand this, let us remind ourselves of how the Berry phase is computed.

\begin{figure}
\includegraphics[width=0.8\columnwidth,angle=0]{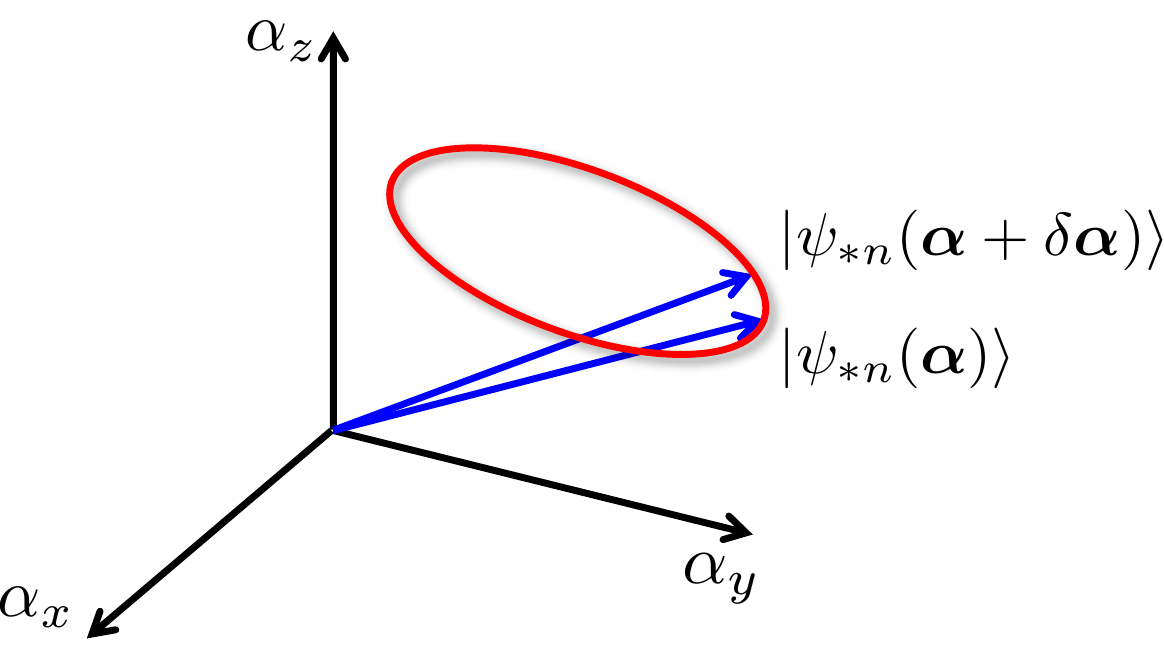}
\caption{Schematic diagram showing how the Berry phase is computed.
}
\label{fig:Berry_phase}  
\end{figure}   

The concept of the Berry phase was originally introduced in the context of the adiabatic approximation~\cite{Berry1984}.
The adiabatic approximation concerns the situation, where the system is subject to a very slowly varying perturbation in time.
Specifically, in the adiabatic approximation, the solution of the full time-dependent Schr\"{o}dinger equation is approximated to be proportional to the instantaneous eigenstate of the Hamiltonian as if time is just a fixed parameter.
Mathematically,
\begin{align}
|\psi_n(t)\rangle \simeq |\psi_{*n}(t)\rangle e^{-\frac{i}{\hbar}\int^t_0 dt^\prime E_{*n}(t^\prime)} e^{i\gamma_n(t)} ,
\end{align}
where $|\psi_{*n}(t)\rangle$ is the $n$-th instantaneous eigenstate with the instantaneous eigenvalue $E_{*n}(t)$ at a given time $t$.
Here, $\gamma_n(t)$ is the Berry phase of the $n$-th instantaneous eigenstate.
Actually, the existence of this additional phase had been known even before Berry. 
However, it was always assumed that this phase could be eliminated by devising an appropriate gauge transformation.
As realized by Berry, this is not always possible. 
Sometimes, there emerges a situation, where the additional phase factor survives and generates a physically observable consequence.

To see this, it is instructive to consider a concrete example, where the Hamiltonian depends on a 3D vector parameter $\boldsymbol{\alpha}$, which is in turn a very slowly varying function of time.
In this situation, let us compute how the additional phase $\gamma_n$ depends on the $n$-th instantaneous eigenstate $|\psi_{*n}(\boldsymbol{\alpha})\rangle$.
Specifically, we begin by computing the overlap between the $n$-th instantaneous eigenstates at $\boldsymbol{\alpha}$ and $\boldsymbol{\alpha}+\delta\boldsymbol{\alpha}$:
\begin{align}
\langle\psi_{*n}(\boldsymbol{\alpha}) |\psi_{*n}(\boldsymbol{\alpha}+ \delta\boldsymbol{\alpha})\rangle
&\simeq 1+ \delta\boldsymbol{\alpha} \cdot \langle\psi_{*n}(\boldsymbol{\alpha})| \nabla_{\boldsymbol{\alpha}} |\psi_{*n}(\boldsymbol{\alpha})\rangle
\nonumber \\
&\simeq e^{-i \delta\boldsymbol{\alpha} \cdot \mathbfcal{A}_n(\boldsymbol{\alpha})} ,
\label{eq:Overlap_Berry_connection}
\end{align}
where 
\begin{align}
\mathbfcal{A}_n(\boldsymbol{\alpha})=i\langle\psi_{*n}(\boldsymbol{\alpha})| \nabla_{\boldsymbol{\alpha}} |\psi_{*n}(\boldsymbol{\alpha})\rangle .
\label{eq:Berry_connection}
\end{align}
See Fig.~\ref{fig:Berry_phase} for a schematic diagram. 
The last expression in Eq.~\eqref{eq:Overlap_Berry_connection} looks as if it is the Aharonov-Bohm phase in the parameter space induced by an effective vector potential $\mathbfcal{A}_n(\boldsymbol{\alpha})$. 
Actually, it turns out that it can be regarded as being such. 
For the time being, let us assume so and proceed further.
Then, the effective magnetic field can be defined by taking the curl of $\mathbfcal{A}_n(\boldsymbol{\alpha})$:
\begin{align}
\mathbfcal{B}_n(\boldsymbol{\alpha}) &= \nabla_{\boldsymbol{\alpha}} \times \mathbfcal{A}_n(\boldsymbol{\alpha})
\nonumber \\
&=i \langle \nabla_{\boldsymbol{\alpha}}  \psi_{*n}(\boldsymbol{\alpha})| \times |\nabla_{\boldsymbol{\alpha}} \psi_{*n}(\boldsymbol{\alpha})\rangle ,
\label{eq:Berry_curvature}
\end{align}
which in turn can be used to compute the effective Aharanov-Bohm phase:
\begin{align}
\gamma_n = i \int_A d{\bf S} \cdot \langle \nabla_{\boldsymbol{\alpha}}  \psi_{*n}(\boldsymbol{\alpha})| \times |\nabla_{\boldsymbol{\alpha}} \psi_{*n}(\boldsymbol{\alpha})\rangle ,
\label{eq:Berry_flux}
\end{align}
where $A$ denotes an area in the parameter space bounded by a closed boundary, or path. 
Technically, $\mathbfcal{A}_n(\boldsymbol{\alpha})$, $\mathbfcal{B}_n(\boldsymbol{\alpha})$, and $\gamma_n$ are called the Berry connection, curvature, and flux, respectively.

Now, it is important to notice the similarity between Eqs.~\eqref{eq:Chern_number} and \eqref{eq:Berry_flux}.
The two expressions are essentially identical with $\boldsymbol{\alpha}$ and ${\bf k}$ playing the corresponding roles.
In Eq.~\eqref{eq:Chern_number}, the area, $A$, is simply equal to the entire Brillouin zone.
The similarity between the two expressions is very important not only because it triggered the conceptual development eventually leading to topological matter, but also because it provides a natural explanation for the reason why the Chern number should be quantized.

The reason is fundamentally due to the single-valuedness of the wave function imposing a strict condition for the magnetic charge of the Dirac monopole, $q_m$, which is present in the Hamiltonian parameter space.
As first discovered by Dirac himself, $q_m$ should be an integer multiple of the flux quantum divided by the solid angle, i.e., $4\pi$ in order for the electron wave function to be single valued. 
Actually, the original formulation was given in such a form that the product between the electron charge $e$ and the magnetic charge $q_m$ should be quantized as follows (in the Gaussian units):
\begin{align}
\frac{e q_m}{\hbar c/2} \in \mathbb{Z} ,
\end{align}
meaning that 
$4 \pi q_m/\phi_0 \in \mathbb{Z}$.
As shown in one of the following sections, the archetypal Hamiltonian of the 2D Chern insulator has the Dirac monopole in the Hamiltonian parameter space, whose magnetic charge is $q_m=\pm \frac{\phi_0}{4\pi}$ with the sign depending on the energy level.

{\bf Hofstadter's butterfly.}
There is a subtle, but very intriguing problem that arises when one tries to apply the TKNN formula to the IQHS as it is written. 
The problem is that the TKNN formula requires the momentum to be a good quantum number.
Unfortunately, the presence of the magnetic field in the IQHS breaks the translational symmetry, at least, in the Hamiltonian level since the vector potential depends on position.
A solution to this problem is to introduce an additional periodic potential forming the lattice structure and enlarge the unit cell to enclose an integer number of flux quanta, called the magnetic unit cell.

\begin{figure}
\includegraphics[width=0.8\columnwidth,angle=0]{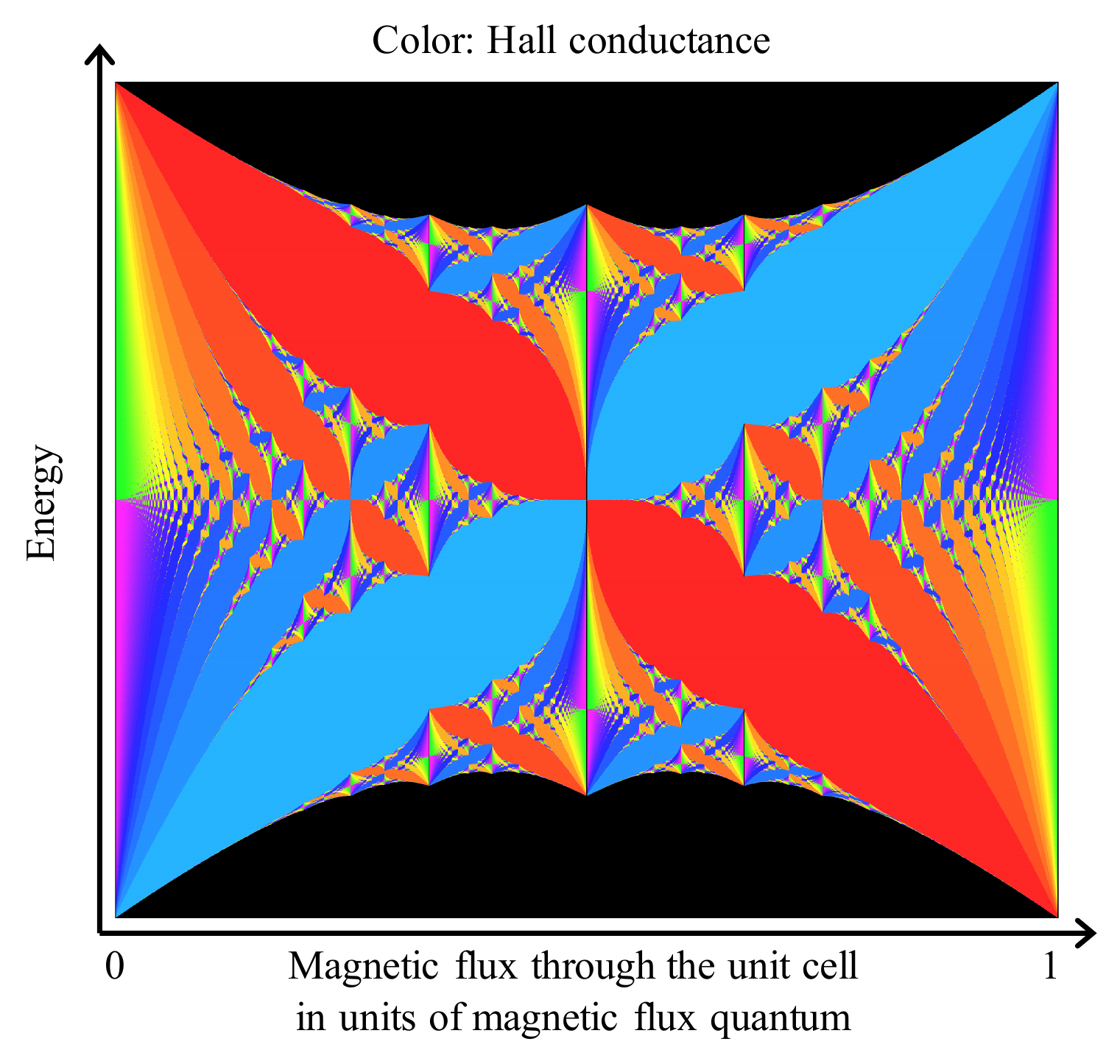}
\caption{Hofstadter's butterfly in the square lattice as a function of magnetic flux through the unit cell.
Belonging to the public domain, the main figure is taken from wikipedia.
The axis labels are added by the current author of this review.
Note that different colors indicate different quantized values of the Chern number, or the Hall conductance.
}
\label{fig:Hofstadter_butterfly}  
\end{figure}

Surprisingly, it is found in this situation that the energy levels exhibit a fractal structure known as Hofstadter's butterfly~\cite{Hofstadter1976}.
See Fig.~\ref{fig:Hofstadter_butterfly} for Hofstadter's butterfly in the square lattice.
It is important to note that the complex energy levels reduce to the usual Landau levels in the limit of the magnetic flux approaching either 0 or 1, or the continuum limit. 
Such a limiting process provides a guarantee that, while not easy to compute directly, the Chern number of the lowest Landau level (for that matter, any Landau levels) can be regarded as being quantized as unity.

{\bf Haldane model: Chern insulator}

Setting aside the issue of Hofstadter's butterfly, the TKNN formula stimulated the imaginations of many researchers including Haldane, who conducted a theoretical study to investigate the possibility of the quantized Hall conductance without Landau levels~\cite{Haldane1988}.
Realizing that the TKNN formula does not require the existence of a finite magnetic field, Haldane constructed a tight-binding model Hamiltonian in graphene with both nearest and next-nearest neighbor hopping in the presence of an appropriately modulating, but net-zero magnetic field.
It turns out that the model constructed by Haldane, called the Haldane model from this forward, is an exact embodiment of the 1D idea mentioned in the beginning of this review for the 2D generalization.

\begin{figure*}
\begin{center}
\includegraphics[width=1.4\columnwidth,angle=0]{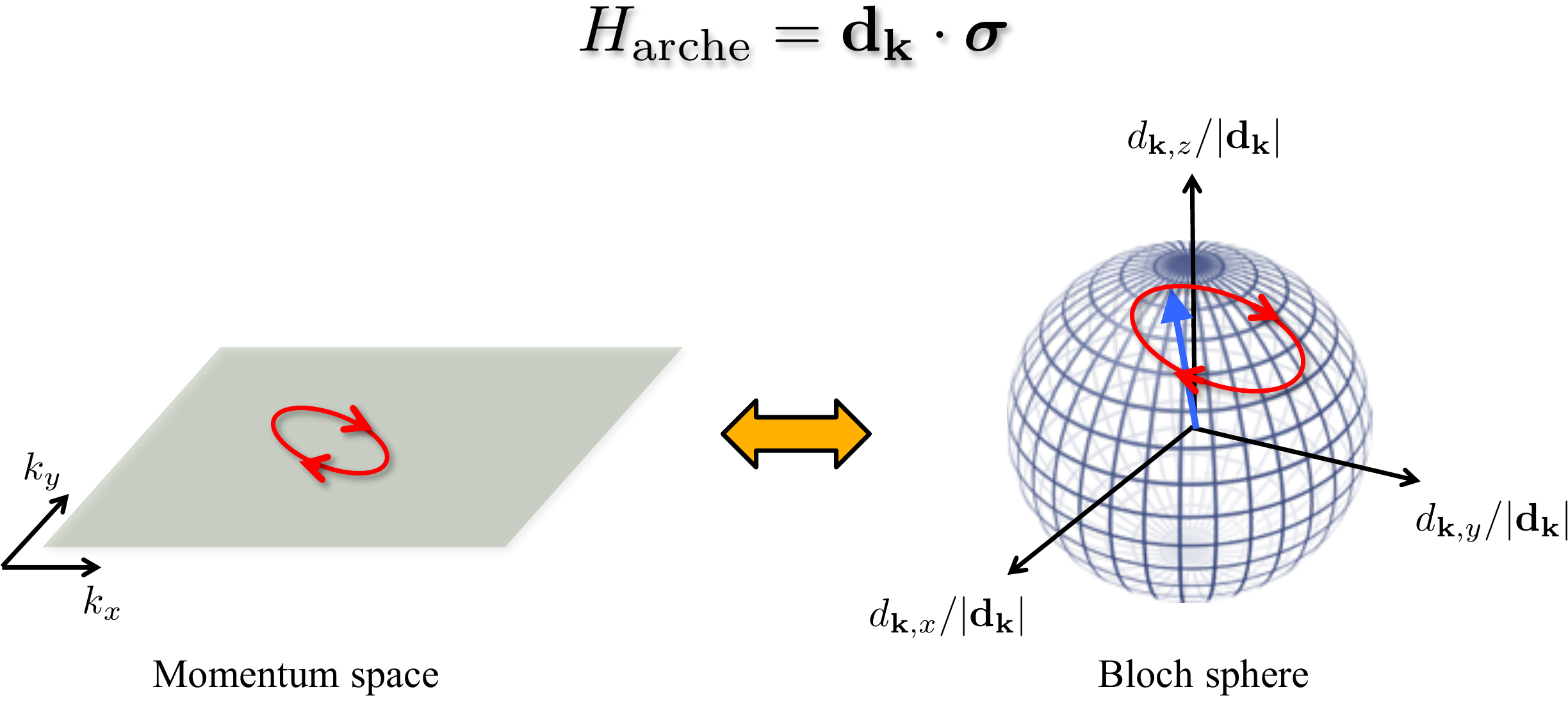}
\caption{Mapping between the 2D Brillouin zone and the surface of the Bloch sphere. 
It is important to note that the archetypal Hamiltonian of the 2D Chern insulator, $H_{\rm arche}$, can be regarded as a mapping between these two parameter spaces. 
}
\label{fig:Mapping_momentum_Bloch} 
\end{center} 
\end{figure*}   

The Haldane model can be regarded as a mapping from the 2D Brillouin zone to the surface of the Bloch sphere.
That is, a momentum eigenstate in the 2D Brillouin zone can be mapped onto a unit vector, whose end point is located at a position in the surface of the Bloch sphere. 
The Bloch sphere contains a Dirac monopole with the monopole strength being $\pm 1/2$, provided that the Hamiltonian is given as the following archetypal form:
\begin{align}
H_{\rm arche}={\bf d}_{\bf k} \cdot \boldsymbol{\sigma}
\label{eq:H_arche}
\end{align} 
where $\boldsymbol{\sigma}$ is the Pauli matrix vector.
The momentum dependence of the Hamiltonian parameter, ${\bf d}_{\bf k}$, determines whether the model is topologically trivial or not.
Specifically, if the normalized vector $\hat{\bf d}_{\bf k}={\bf d}_{\bf k}/|{\bf d}_{\bf k}|$ wraps around the Bloch sphere entirely (or, any integer number of times), the total Berry flux becomes non-trivial due to the presence of the Dirac monopole at the center of the Bloch sphere, i.e., at ${\bf d}_{\bf k}=0$.  
See Fig.~\ref{fig:Mapping_momentum_Bloch} for a schematic diagram.

Intriguingly, the Hamiltonian in Eq.~\eqref{eq:H_arche} is exactly the same Hamiltonian describing the Rabi oscillation in the presence of time-periodic magnetic field. 
The Rabi oscillation is mostly known for the magnetic resonance.
However, the Hamiltonian describing the Rabi oscillation has exactly the same monopole structure in its parameter space as the 2D topological insulator.
Thus, a complete understanding of the Rabi oscillation should be sufficient for that of the 2D topological insulator, or, strictly speaking, 2D Chern insulator. 

To be specific, let us write the Hamiltonian for the Rabi oscillation:
\begin{align}
H_{\rm Rabi} = \boldsymbol{\alpha} \cdot \boldsymbol{\sigma} ,
\label{eq:H_Rabi}
\end{align}
where $\boldsymbol{\alpha}$ is essentially equal to the rotating magnetic field in the actual Rabi oscillation problem. 
Obviously, $\boldsymbol{\alpha}$ corresponds to ${\bf d}_{\bf k}$ in Eq.~\eqref{eq:H_arche}.
In the actual Rabi oscillation problem, $\boldsymbol{\alpha}=\mu [ B_1\cos{(\Omega t)}\hat{x} +B_1\sin{(\Omega t)}\hat{y} +B_0\hat{z} ]$ with $\mu$ being the magnetic moment and $\Omega$ being the driving frequency.
The specific form of the time dependence is not important here since we are interested in the Berry phase only.
Now, we would like to compute the Berry curvature of this Rabi Hamiltonian to show that there is a Dirac monopole at $\boldsymbol{\alpha}=0$.

It is shown in Eq.~\eqref{eq:Berry_curvature} that the Berry curvature can be written as follows:
\begin{align}
\mathbfcal{B}_{\pm}(\boldsymbol{\alpha}) &=i \langle \nabla_{\boldsymbol{\alpha}}  \psi_{\pm}(\boldsymbol{\alpha})| \times |\nabla_{\boldsymbol{\alpha}} \psi_{\pm}(\boldsymbol{\alpha})\rangle ,
\label{eq:Berry_curvature2}
\end{align}  
where $|\psi_{\pm}(\boldsymbol{\alpha})\rangle$ denotes the eigenstate of the Rabi Hamiltonian in Eq.~\eqref{eq:H_Rabi} with energy eigenvalue $\epsilon_{\pm}(\boldsymbol{\alpha})=\pm |\boldsymbol{\alpha}|$.
Actually, the expression in Eq.~\eqref{eq:Berry_curvature2} is rather inconvenient since one has to take the derivatives of the eigenstate. 
Instead, one can rewrite Eq.~\eqref{eq:Berry_curvature2} in a more convenient that does not involve any derivatives of the eigenstate:
\begin{widetext}
\begin{align}
\mathbfcal{B}_n(\boldsymbol{\alpha}) &=i \langle \nabla_{\boldsymbol{\alpha}}  \psi_n(\boldsymbol{\alpha})| \times |\nabla_{\boldsymbol{\alpha}} \psi_n(\boldsymbol{\alpha})\rangle ,
\nonumber \\
&= i \sum_{m \neq n} \langle \nabla_{\boldsymbol{\alpha}}  \psi_n(\boldsymbol{\alpha}) | \psi_m(\boldsymbol{\alpha}) \rangle \times \langle \psi_m(\boldsymbol{\alpha})  |\nabla_{\boldsymbol{\alpha}} \psi_n(\boldsymbol{\alpha})\rangle ,
\nonumber \\
&= i \sum_{m \neq n}
\frac{
\langle  \psi_n(\boldsymbol{\alpha}) |\nabla_{\boldsymbol{\alpha}}H(\boldsymbol{\alpha})| \psi_m(\boldsymbol{\alpha}) \rangle \times \langle \psi_m(\boldsymbol{\alpha})  |\nabla_{\boldsymbol{\alpha}} H(\boldsymbol{\alpha}) | \psi_n(\boldsymbol{\alpha})\rangle
}{[\epsilon_n(\boldsymbol{\alpha})-\epsilon_m(\boldsymbol{\alpha})]^2} ,
\label{eq:Berry_curvature3}
\end{align}
\end{widetext}
where $n$ and $m$ are the level indices denoting $\pm$ for $|\psi_{\pm}(\boldsymbol{\alpha})\rangle$.
The second line in the above equation is obtained owing to the completeness of the Hamiltonian eigenstates.
The third line can be obtained by multiplying $\langle \psi_m(\boldsymbol{\alpha}) | \nabla_{\boldsymbol{\alpha}}$ to both sides of the eigenvalue equation:
\begin{align}
\langle \psi_m(\boldsymbol{\alpha})  |\nabla_{\boldsymbol{\alpha}} \psi_n(\boldsymbol{\alpha})\rangle 
=\frac{\langle \psi_m(\boldsymbol{\alpha})  |\nabla_{\boldsymbol{\alpha}} H(\boldsymbol{\alpha}) | \psi_n(\boldsymbol{\alpha})\rangle
}{\epsilon_n(\boldsymbol{\alpha})-\epsilon_m(\boldsymbol{\alpha})} .
\end{align}
Note that the final expression in Eq.~\eqref{eq:Berry_curvature3} involves only the derivatives of the Hamiltonian, not the eigenstates.

By using Eq.~\eqref{eq:Berry_curvature3}, one can show that
\begin{align}
\mathbfcal{B}_{\pm}(\boldsymbol{\alpha}) &= \mp \frac{1}{2} \frac{\hat{\boldsymbol{\alpha}}}{\alpha^2} ,
\end{align}
which is identical to the usual inverse-square law of the electric field induced by a point electric charge.
This means that there is a Dirac monopole with the magnetic charge equal to $\pm 1/2$ at $\boldsymbol{\alpha}=0$.

Due to the identical form of the Hamiltonian, a Dirac monopole should also exist in the archetypal Hamiltonian $H_{\rm arche}$ in Eq.~\eqref{eq:H_arche}.
Consequently, if the normalized vector $\hat{\bf d}_{\bf k}$ wraps around the Bloch sphere entirely, the total Berry flux is given as the product between the magnetic charge of the Dirac monopole, $\pm 1/2$, and the solid angle, $4\pi$, amounting to $\pm 2\pi$. 
Since the Chern number is defined as the total Berry flux divided by $2\pi$, this means that the Chern number becomes simply $\pm 1$ for the topologically non-trivial state of $H_{\rm arche}$.

Concretely, the Hamiltonian for the Haldane model has the following mathematical form~\cite{Haldane1988}:
\begin{align}
H_{\rm Haldane} = 
\left(
\begin{array}{cc}
g_{+,{\bf k}} & f^*_{\bf k} \\
f_{\bf k} & g_{-,{\bf k}} \\
\end{array}
\right) ,
\label{eq:H_Haldane}
\end{align}
where 
\begin{align}
f_{\bf k} &= t_1 \sum_i e^{i{\bf k}\cdot{\bf a}_i}  , \\
g_{\pm,{\bf k}} &= \pm \Delta+2 t_2 \sum_i \cos{({\bf k}\cdot{\bf b}_i \pm \phi)} ,
\end{align}
where $t_1$ and $t_2$ are the hopping parameters between nearest and next nearest neighbors, respectively.
Similarly, ${\bf a}_i$ and ${\bf b}_i$ are the displacement vectors connecting between nearest and next nearest neighbors, respectively.
$\Delta$ is the on-site energy difference between the sites at sublattice $A$ and $B$.
$\phi$ is an appropriate phase acquired by the hopping between next nearest neighbors due to a modulating, but net-zero magnetic field inside the hexagonal unit cell.

The Hamiltonian in Eq.~\eqref{eq:H_Haldane} can be rewritten in the form of Eq.~\eqref{eq:H_arche}:
\begin{align}
H_{\rm Haldane} = h_{\bf k} {\bf I} +{\bf d}_{\bf k}\cdot\boldsymbol{\sigma} 
\label{eq:H_Haldane2}
\end{align}
where 
\begin{align}
h_{\bf k} &= 2 t_2 \cos{\phi}\sum_i \cos{({\bf k}\cdot{\bf b}_i)} , \\
d_{{\bf k},x} &= {\rm Re} f_{\bf k} = t_1 \sum_i \cos{({\bf k}\cdot{\bf a}_i)} , \\
d_{{\bf k},y} &= {\rm Im} f_{\bf k} = t_1 \sum_i \sin{({\bf k}\cdot{\bf a}_i)} , \\
d_{{\bf k},z} &= \Delta -2 t_2 \sin{\phi}\sum_i \sin{({\bf k}\cdot{\bf b}_i)} .
\end{align} 
Here, note that, while depending on ${\bf k}$, the $h_{\bf k} {\bf I}$ term shifts both energies of the conduction and valence bands together so that the direct gap between the two bands remains the same even if we ignore it.
Therefore, the $h_{\bf k} {\bf I}$ term can be ignored unless the two bands overlap in different momenta; that is, the band gap closes indirectly.

To test whether the Haldane model is topologically trivial or non-trivial at a particular choice of the parameters, $t_1$, $t_2$, $\Delta$, and $\phi$, it is convenient to expand $H_{\rm Haldane}$ near the Dirac points.
It is important to note that the Haldane model reduces to the usual tight-binding model of graphene with nearest-neighbor hopping only when $t_2=\Delta=0$.
In this situation, the gap closes at the usual Dirac points at ${\bf K}$ and ${\bf K}^\prime$.
With addition of non-zero $t_2$ and $\Delta$, the gap opens up, but its magnitude remains to be the minimum at the Dirac points.
Considering that the magnitude of the gap indicates the distance from the Dirac monopole in the Hamiltonian parameter space, the minimum gap position plays a crucial role in determining the topology, or the wrapping of $\hat{\bf d}_{\bf k}$ around the Bloch sphere.
Specifically, ${\bf d}_{\bf k}$ can be expanded near the Dirac points as follows:
\begin{align}
d_{{\bf q},x} &\simeq  A q_x, \label{eq:d_qx_expand}  \\
d_{{\bf q},y} &\simeq  \pm A q_y, \label{eq:d_qy_expand} \\
d_{{\bf q},z} &\simeq M+B(q^2_x+q^2_y) , \label{eq:d_qz_expand} 
\end{align} 
with $A$, $B$, and $M$ depending on $t_1$, $t_2$, $\Delta$, and $\phi$.
Specifically,
\begin{align}
A &=  \frac{3}{2} t_1 , \\
B &=  \pm \frac{9\sqrt{3}}{4} t_2 \sin{\phi} , \\
M &= \Delta \mp 3\sqrt{3} t_2 \sin{\phi} ,
\end{align} 
where we have ignored an unimportant phase factor of $A$.
Above, ${\bf q}=(q_x,q_y)$ denotes the displacement vector measured from the Dirac points; that is, ${\bf k}={\bf K}+{\bf q}$ or ${\bf k}={\bf K}^\prime+{\bf q}$. 
Note that the sign in the right-hand side of Eq.~\eqref{eq:d_qy_expand} depends on near which Dirac point, ${\bf K}$ or ${\bf K^\prime}$, ${\bf d}_{\bf k}$ is expanded.

\begin{figure*}
\begin{center}
\includegraphics[width=1.4\columnwidth,angle=0]{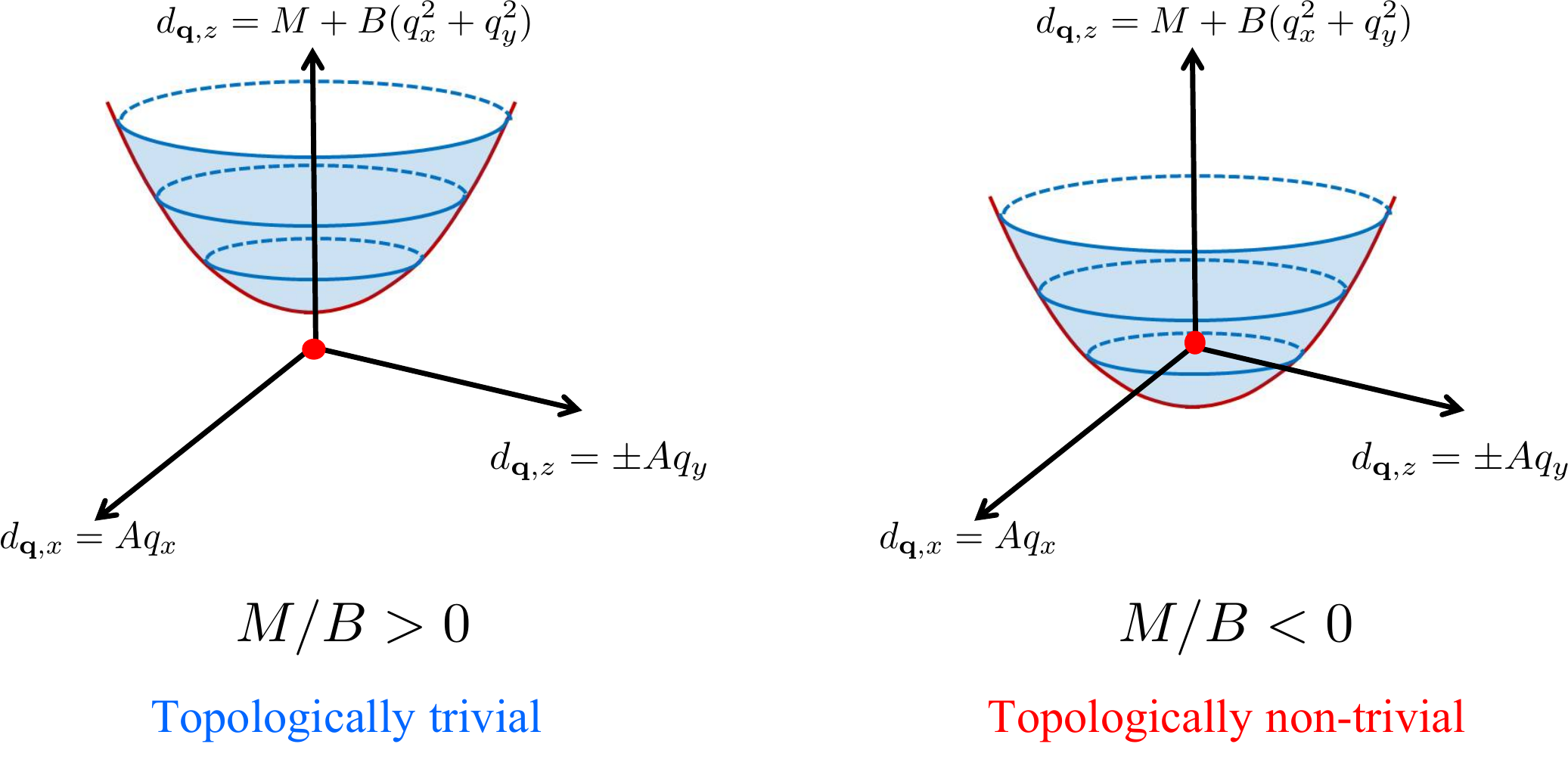}
\caption{Schematic diagram illustrating the topology test of the Haldane model, or generally the archetypal model Hamiltonian of the 2D Chern insulator. 
It is important to note that, generically, the Hamiltonian parameter ${\bf d}_{\bf k}=(d_{{\bf q},x},d_{{\bf q},y},d_{{\bf q},z})$ can be always expanded as $(A q_x, \pm A q_y, M+B(q_x^2+q_y^2))$ near the momentum point, where the band gap is minimum.
As explained in the main text, the topology is non-trivial if $M/B<0$ and trivial otherwise.
}
\label{fig:Topology_test} 
\end{center} 
\end{figure*}

Then, the condition for a complete wrapping of the unit vector $\hat{\bf d}_{\bf k}$ around the Bloch sphere can be visualized as the condition that the parabola sheet formed by ${\bf d}_{\bf k}$ (not the unit vector $\hat{\bf d}_{\bf k}$) encloses a Dirac monopole at the origin.
This condition is simply determined by the sign of $M/B$.
That is, the topology is non-trivial if $M/B<0$ and trivial otherwise.
See Fig.~\ref{fig:Topology_test} for a schematic diagram.
Note that $A$ is not important as far as the topology is concerned.
For the Haldane model, this condition amounts to $|\Delta/t_2|<3\sqrt{3}|\sin{\phi}|$, which is exactly the same formula obtained by Haldane in Ref.~\cite{Haldane1988}.

{\bf Topological insulator}

In a narrow sense, the term ``topological insulator'' indicates an appropriate time-reversal invariant version of the Chern insulator with two spin species.
As mentioned previously, in 2D, this means two independent copies of the Chern insulator with opposite Chern numbers for different spins, preserving the time-reversal symmetry as a whole.
It was perhaps the existence of this time-reversal invariant topological insulator in 2D that sparked a remarkable attention from the entire community of condensed matter physics.
While there have been many prior attempts, such an existence was first realized by Kane and Mele in their seminal paper~\cite{Kane2005}
in the form recognized as being complete in the modern standard.
For this reason, let us begin our discussion on the 2D topological insulator with the model Hamiltonian proposed by Kane and Mele.

The microscopic model proposed by Kane and Mele is based on the spin-orbit-coupled graphene, which reduces to two copies of the Haldane model at an appropriate parameter.
Concretely, the Kane-Mele Hamiltonian is written as follows:
\begin{align}
H_{\rm Kane\mbox{-}Mele} = &-t \sum_{\langle i,j \rangle, \sigma} ( c^\dagger_{i\sigma} c_{j\sigma} + {\rm H.c.} )
\nonumber \\
&+i \lambda_{\rm SO} \sum_{\langle\langle i,j \rangle\rangle, \sigma} \sigma \nu_{ij} ( c^\dagger_{i\sigma} c_{j\sigma} + {\rm H.c.} ) ,
\label{H_KM}
\end{align}
where $\sigma=\pm 1$ for spin up and down, respectively. 
Also, defined between next nearest neighbors, $\nu_{ij}=1$ $(-1)$ if the hopping from ${\bf r}_i$ to ${\bf r}_j$ indicates a counterclockwise (clockwise) hopping with respect to the center of the hexagonal unit cell.
In the momentum space representation, the Kane-Mele Hamiltonian can be written in the archetypal form similar to that of the Haldane Hamiltonian:
\begin{align}
H^{(\uparrow)}_{\rm Kane\mbox{-}Mele}({\bf k}) &= {\bf d}_{\bf k}\cdot\boldsymbol{\sigma}  
\label{eq:H_KM2_up} \\
H^{(\downarrow)}_{\rm Kane\mbox{-}Mele}({\bf k}) &= [H^{(\uparrow)}_{\rm Kane\mbox{-}Mele}(-{\bf k})]^* ,
\label{eq:H_KM2_down}
\end{align}
where $\uparrow$ and $\downarrow$ indicate the up and down spins, respectively, and
\begin{align}
d_{{\bf k},x} &= t \sum_i \cos{({\bf k}\cdot{\bf a}_i)} , \\
d_{{\bf k},y} &= t \sum_i \sin{({\bf k}\cdot{\bf a}_i)} , \\
d_{{\bf k},z} &= 2 \lambda_{\rm SO}\sum_i \sin{({\bf k}\cdot{\bf b}_i)} .
\end{align} 
Equations~\eqref{eq:H_KM2_up} and \eqref{eq:H_KM2_down} indicate that, forming the Chern insulators individually, the two Hamiltonian components for different spin species are time-reversal conjugate to each other, making the whole system time-reversal invariant. 
Similar to the Haldane model, $d_{{\bf k},z}$ can be expanded near the Dirac points:
\begin{align}
d_{{\bf q},z}/\lambda_{\rm SO} &\simeq \pm3\sqrt{3} \mp \frac{9\sqrt{3}}{4} (q^2_x+q^2_y) , 
\label{eq:d_qz_expand_KM}
\end{align} 
where, again, the sign depends on near which Dirac point $d_{{\bf q},z}$ is expanded.
Equation~\eqref{eq:d_qz_expand_KM} shows that the topology is always non-trivial for the Kane-Mele model.

{\bf Generalization to 3D topological insulator.}
Na\"{i}vely, it may be expected that the 3D topological insulator can be also constructed by generalizing the 2D concept to a compact 3D manifold with appropriate Dirac monopoles.
Unfortunately, this does not work.
Mathematically, it is not possible to generalize the Chern number to 3D.
Fortunately, however, if both inversion and time-reversal symmetries are present, the 3D band topology can be characterized by four $\mathbb{Z}_2$ invariants, $(\nu_0;\nu_1,\nu_2,\nu_3)$, which depend on the parities of the time-reversal operator, $\delta_{i (=1,\cdots,8)},$  at eight time reversal invariant momentum (TRIM) points~\cite{Fu2007, Fu2007b}.

To provide an intuition for the $\mathbb{Z}_2$ invariants, let us begin with a simple example of the 3D topological insulator, which is a stack of 2D topological insulators. 
This is called the {\it weak} 3D topological insulator. 
The reason why it is called being weak is that its edge states are protected only along a certain direction.

The so-called strong 3D topological insulator, whose edge states are protected in all directions, is achieved by relaxing the idea that the Chern number can be somehow generalized to 3D by using the topological information of the entire 3D Brillouin zone. 
In fact, there is an immediate problem if one tries to define the Chern number in 3D, where the spin degree of freedom is generally coupled with the orbital counterpart, and therefore the Chern number cannot be defined for each spin species separately.   
A literal, but effective solution to this problem is to define the Chern number only when it can be done. 
In the presence of both inversion and time-reversal symmetries, there are certain 2D planes in the 3D Brillouin zone, where the Chern number can be defined separately for each spin species. 
These 2D planes are none other than those containing the TRIM points~\cite{Lee2015}.
For convenience, let us call such 2D planes the TRIM planes.

To concretely show how this can be done, let us consider a generic form of the Hamiltonian for 3D topological insulator, which can be expanded near the minimum gap position, e.g. the $\Gamma$ point for BiSe-family materials~\cite{Hasan2010, Qi2011_Review}: 
\begin{widetext}
\begin{align}
H_\textrm{3D}(\mathbf{k})
= \epsilon_\mathbf{k} \mathbb{I}_4 
+ \left(
\begin{array}{cccc}
M +B_1 k_\perp^2 +B_2 k_z^2 & A_1(k_x+ik_y) & 0 & A_2 k_z \\
A_1(k_x-ik_y) & -(M +B_1 k_\perp^2 +B_2 k_z^2) & A_2 k_z & 0 \\
0 & A_2 k_z & M +B_1 k_\perp^2 +B_2 k_z^2 & -A_1(k_x-ik_y) \\
A_2 k_z & 0 & -A_1(k_x+ik_y) & -(M +B_1 k_\perp^2 +B_2 k_z^2)
\end{array}
\right),
\label{eq:H_3D}
\end{align}
\end{widetext}
where $\mathbb{I}_4$ is the $4\times 4$ identity matrix, and the overall energy shift $\epsilon_{\bf k}$ can be expanded as $C + D_1 k_\perp^2 +D_2 k_z^2$ with $k_\perp^2=k_x^2+k_y^2$.
While the original microscopic Hamiltonian can be very complicated, the essential properties of the 3D topological insulator can be well captured by the above expanded Hamiltonian or its minimally lattice-regularized version constructed via $k_i \rightarrow \sin{k_i}$ and $k_i^2 \rightarrow 2(1-\cos{k_i})$ with $i=x, y, z$.

Let us begin by investigating what happens at $k_z=0$, one of the TRIM planes.
In this situation, $H_{\rm 3D}({\bf k})$ reduces to a $2\times 2$ block-diagonalized matrix describing two independent copies of the Chern insulator with opposite Chern numbers for different spins just like the 2D topological insulator.   
Therefore, at least at $k_z=0$, each spin species can have the well-defined Chern number with an opposite value to each other.
Also, due to the reason explained in the preceding section, the topology is determined by the sign of $M/B_1$.

Next, we investigate what happens at $k_z=\pi$, another one of the TRIM planes.
In this situation, one should consider the lattice-regularized Hamiltonian with $k_z \rightarrow \sin{k_z}|_{k_z=\pi}=0$ and $k_z^2 \rightarrow 2(1-\cos{k_z})|_{k_z=\pi}=4$. 
Again, here, $H_{\rm 3D}({\bf k})$ reduces to a $2\times 2$ block-diagonalized matrix describing two independent copies of the Chern insulator. 
A difference is that the topology is now determined by the sign of $(M+4B_2)/B_1$.

Now, we arrive at the stage, where the strong 3D topological insulator can be defined. 
The strong 3D topological insulator can be defined as such a topological insulator that its 2D topology is non-trivial at $k_z=0$ while trivial at $k_z=\pi$, or vice versa.
While it is not easy to see at this stage, one can show that the definition for the strong 3D topological insulator does not depend on the choice of the axis~\cite{Lee2015}.
That is, one can choose the TRIM planes at $k_x=0$ and $k_x=\pi$, or those at  $k_y=0$ and $k_y=\pi$ instead of those at $k_z=0$ and $k_z=\pi$.

Actually, there is an equivalent, but much more convenient way of tracking the 3D topology instead of monitoring the 2D topology of the TRIM planes.
That is the above-mentioned $\mathbb{Z}_2$ invariants.
Below, I discuss how the 2D topology of the TRIM planes can be connected with the $\mathbb{Z}_2$ invariants.

In the presence of the time-reversal symmetry, the Chern numbers are always opposite between different spins, meaning that the 2D band topology is fully characterized by the Chern number difference.
Motivated by the analogy between the charge and time-reversal polarization, the Chern number difference can be alternatively computed in a discrete form, which is formulated in terms of the parities of the time-reversal operator, $\delta_{i(=1,2,3,4)}$, at four TRIM points~\cite{Kane2005a}:
\begin{align}
(-1)^{\nu_{\rm 2D}} = \prod_{i=1}^4 \delta_i .
\label{eq:2D_Z2_invariant}
\end{align}
It is important to note that the above 2D $\mathbb{Z}_2$ invariant, $\nu_{\rm 2D}$, is exactly identical to half the Chern number difference computed via the integral form in Eq.~\eqref{eq:Chern_number}: 
\begin{align}
\nu_{\rm 2D} = \frac{\mathcal{C}_\uparrow - \mathcal{C}_\downarrow}{2}~~\textrm{(mod 2)} ,
\label{eq:2D_Z2_invariant2}
\end{align}
where $\mathcal{C}_\uparrow$ and $\mathcal{C}_\downarrow$ are the Chern numbers for the up and down spins, respectively
As shown below, the fact that the 2D topological invariant can be computed in a discrete form plays an important role in defining the 3D topological invariants.

Na\"{i}vely, since there are generally six TRIM planes in 3D, there could be the same number of $\mathbb{Z}_2$ topological invariants. 
For example, in the cubic lattice, the TRIM planes are those defined by $k_x=0$ or $\pi$ and the others with $k_x$ replaced by $k_y$ and $k_z$.
It turns out, however, that we only need four distinct numbers $(\nu_0;\nu_1,\nu_2,\nu_3)$ to fully specify the 3D topology.
Note that each of the $\mathbb{Z}_2$ indices, $(\nu_0;\nu_1,\nu_2,\nu_3)$, is the 2D $\mathbb{Z}_2$ invariant for its specific 2D plane computed via either Eq.~\eqref{eq:2D_Z2_invariant} or \eqref{eq:2D_Z2_invariant2}.

Called the weak indices, the three $\mathbb{Z}_2$ topological invariants, $\nu_1$, $\nu_2$, and $\nu_3$, are the usual 2D $\mathbb{Z}_2$ invariants for three TRIM planes, say, defined by $k_x=0$, $k_y=0$, and $k_z=0$, respectively.
That is, $\nu_1=\nu_{k_x=0}= ({\cal C}_{\uparrow,k_x=0} -{\cal C}_{\downarrow,k_x=0})/2$ (mod 2) with $\nu_2$ and $\nu_3$ defined similarly by using the $k_y=0$ and $k_z=0$ TRIM planes, respectively.

Called the strong index, $\nu_0$ determines if a given 3D topological insulator is weak or strong. 
Specifically, $\nu_0$ becomes unity (zero) if the $\mathbb{Z}_2$ topological invariant of a given TRIM plane is different from (the same as) that of the opposing TRIM plane in the same direction.
That is, $\nu_0=1$ and 0 if $\nu_{k_z=0}$ is different from and the same as $\nu_{k_z=\pi}$, respectively.
It is important to remember that this definition is exactly identical to that obtained in our previous discussion by using the explicit form of the 3D Hamiltonian in Eq.~\eqref{eq:H_3D}.
The fact that the strong 3D topological insulator possesses topologically protected edge states has been confirmed via angle-resolved photoemission spectroscopy (ARPES)~\cite{Hasan2010,Qi2011_Review}.

{\bf Topological semimetal}

After the establishment of the 2D and 3D topological insulators, researchers set out to explore new phases of topological matter extending the concept of topological matter beyond the insulating state.
Such new phases of topological matter include Weyl~\cite{Murakami2007, Wan2011} and Dirac~\cite{Wang2012} semimetals, which are generally called topological semimetals.
Unlike the topological insulator, where Dirac monopoles are avoided in the momentum space, Weyl and Dirac semimetals have Dirac monopoles directly in the momentum space as isolated points. 
Mathematically, the Hamiltonian for Weyl semimetal can be expanded near the point where the gap vanishes as follows:
\begin{align}
H_{\rm Weyl} = v_{F} {\bf k}\cdot\boldsymbol{\sigma} ,
\label{eq:H_Weyl}
\end{align}
where $v_{F}$ is called the Fermi velocity.
The difference between Weyl and Dirac semimetals is determined by whether the Dirac monopoles of different spin species occur in separated or coincidental points in the momentum space.   
Once separated in Weyl semimetal, Dirac monopoles are called Weyl nodes.
In some sense, graphene is the 2D version of the Dirac semimetal since the Dirac nodes are located in the same positions in the Brillouin zone regardless of spin.

Weyl semimetal can be realized rather naturally from the 3D topological insulator by breaking either the time-reversal or the inversion symmetry. 
Weyl semimetal has attracted intense interest from the condensed matter community due to the existence of its peculiar surface state property known as the Fermi arc. 
The Fermi arc is the gapless surface excitation mode connecting between two surface-projected Weyl nodes, forming an open-end segment in stark contrast with the Fermi circle for the usual 2D edge states. 
In this sense, Weyl semimetal is a more interesting topological matter than Dirac semimetal.

\begin{figure}
\begin{center}
\includegraphics[width=1\columnwidth,angle=0]{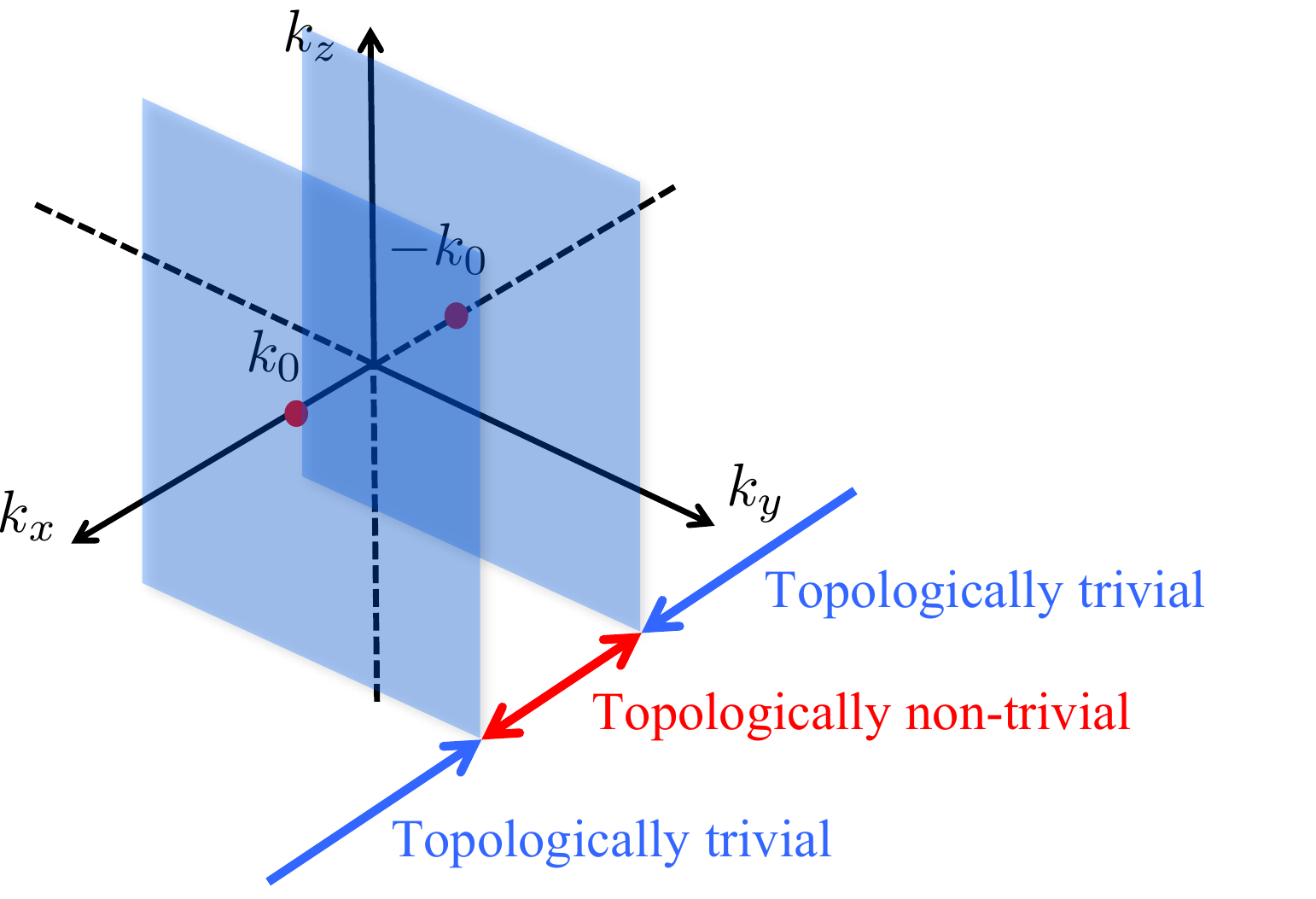}
\caption{Schematic diagram illustrating why there should be a Fermi arc in the surface of Weyl semimetal.
}
\label{fig:Weyl_semimetal} 
\end{center} 
\end{figure}   

To understand why there should be a Fermi arc in the surface of Weyl semimetal, it is convenient to study a simple model Hamiltonian proposed by  Yang {\it et al.}~\cite{Yang2011} describing a time-reversal symmetry-broken Weyl semimetal:
\begin{align}
H_{\rm YLR} = &\left[ -2t(\cos{k_x}-\cos{k_0}) +m(2-\cos{k_y}-\cos{k_z}) \right] \sigma_x
\nonumber \\
&+2t\sin{k_y} \sigma_y +2t\sin{k_z} \sigma_z ,
\label{eq:H_YLR}
\end{align}
which has two Weyl nodes at ${\bf k}=(\pm k_0,0,0)$.
Assuming that $k_0$ is small, $H_{\rm YLR} $ can be expanded near ${\bf k}=0$ as follows:
\begin{align}
H_{\rm YLR} \simeq \left[ t(k^2_x-k^2_0) +\frac{m}{2}(k^2_y+k^2_z) \right] \sigma_x
+2t k_y \sigma_y +2t k_z \sigma_z ,
\label{eq:H_YLR}
\end{align}
which has essentially the same form as the expanded Haldane Hamiltonian in Eq.~\eqref{eq:d_qx_expand}, \eqref{eq:d_qy_expand}, and \eqref{eq:d_qz_expand} with the roles of $d_{{\bf q},z}$, $d_{{\bf q},x}$, and $d_{{\bf q},y}$ in the Haldane Hamiltonian now played by $d_{{\bf k},x}$, $d_{{\bf k},y}$, and $d_{{\bf k},z}$, respectively.

In this situation, the condition for topological non-triviality, $M/B<0$, is now translated into $|k_x|<|k_0|$ since $M=t(k_x^2-k_0^2)$ and $B=m/2$.
That is, the regime defined by $|k_x|<|k_0|$ is topologically non-trivial.
This means that each of the 2D momentum planes within $|k_x|<|k_0|$ can be regarded as an individual 2D Chern insulator. 
See Fig.~\ref{fig:Weyl_semimetal} for a schematic diagram illustrating the situation.

In this situation, those individual 2D Chern insulators have their own chiral edge states.
If so, there must be a line segment of topologically protected chiral edge states existing within the projected surface of the topologically non-trivial regime. 
Such a line segment is the Fermi arc.

{\bf Fractional Chern and topological insulator}

The next challenge in the research of topological matter concerns what happens to topological matter in the presence of strong correlation between electrons.
No matter how complicated their band structure may be, all the topological matters discussed so far can be described by the noninteracting wave function composed of a single Slater determinant.

In the case of topological insulators, the noninteracting wave function can provide a reasonably good state so long as the electron-electron interaction is weaker than the band gap. 
In this situation, the effects of the electron-electron interaction are taken into account in such a level that they just renormalize the band dispersion.
Meanwhile, in the case of Weyl and Dirac semimetals, the noninteracting wave function is protected by a subtle renormalization group process, via which the electron-electron interaction is more or less screened away~\cite{Yang2014,Isobe2016}.

While noninteracting topological matter can be protected against reasonably strong correlation between electrons, more exciting is the possibility that a novel topological matter is {\it induced} by the interaction between electrons~\cite{Tang2011,Sun2011,Neupert2011_PRL,Sheng2011,Wang2011,Regnault2011,Qi2011,Wu2012}.
 A major inspiration comes from the fractional quantum Hall states (FQHSs). 
In the FQHSs, electrons are all confined in the lowest Landau level, which can be regarded as a completely flat topological band without any dispersion what so ever.
Moreover, with the lowest Landau level only fractionally filled, the electron-electron interaction produces highly nonperturbative correlation effects in the FQHSs.  
A question is if a similar interaction-induced topological state can be obtained at a fractional filling of the (nearly) flat Chern band.
If existent, this state would be called the fractional Chern insulator (FCI).
By the similar token, the fractional topological insulator (FTI) can be defined as the interaction-induced topological insulator at a fractional filling of the (nearly) flat Chern band preserving the time-reversal symmetry.

Below, we provide some details on the FQHS, which are necessary to understand how the FCI can be constructed as a lattice analog of the FQHS.

{\bf Fractional quantum Hall state.}
With the first FQHS discovered at 1/3 filling of the lowest Landau level (LLL), initial efforts were devoted to explain why and how an incompressible state can emerge at filling factor $\nu=1/3$.
Eventually, these questions were answered by Laughlin, who put forward the wave function later named after him, the Laughlin wave function~\cite{Laughlin1983}: 
\begin{align}
\Psi_{\rm Laughlin}= \prod_{i<j} (z_i-z_j)^{2p+1} e^{-\sum_k \frac{|z_k|^2}{4l_B^2}} ,
\label{eq:Laughlin}
\end{align}
where $p$ is an integer, related to the filling factor $\nu$ via $\nu=1/(2p+1)$.
Note that $p=1$ corresponds to $\nu=1/3$.

To understand the motivation for the Laughlin state, it is convenient to use the circular gauge rather than the Landau gauge, which was used in one of the preceding sections.
In the circular gauge, the vector potential is set to be ${\bf A}=\frac{B}{2}(-y,x,0)$.
In this situation, the LLL energy eigenstates can be written as follows:
\begin{align}
\psi_m(z) \propto z^m e^{-\frac{|z|^2}{4l_B^2}} ,
\end{align}
where $z=x+iy$ and $m$ is the eigenvalue of the $z$-component angular momentum, $L_z=\hbar(z\frac{\partial}{\partial z}-\bar{z}\frac{\partial}{\partial \bar{z}})$.
Above, the normalization constant is not explicitly shown.
Note that $\psi_m$ describes the cyclotron motion of an electron, which forms a ring with its expectation value of the radius  being equal to $\sqrt{2m}l_B$.

Now, let us discuss how the Laughlin state can be constructed in terms of these LLL eigenstates in the circular gauge.
Specifically, below, we enumerate each of the major ideas leading to the Laughlin state one by one.

(i) Any many-body wave function confined in the LLL should be written solely in terms of the above LLL eigenstates.
That is, the many-body wave function should be a holomorphic function, i.e., a function of complex variables entirely composed of $z$, not $\bar{z}$:
\begin{widetext}
\begin{align}
\Psi(z_1,z_2,\cdots, z_N) = \sum_{\{m_i\}} {\cal C}_{\{m_i\}} {\cal A} \left[ z_1^{m_1} z_2^{m_2} \cdots z_N^{m_N}\right] e^{-\sum_k \frac{|z_k|^2}{4l_B^2}},
\end{align}
\end{widetext}
where ${\cal A}$ is the antisymmetrization operator.
Note that the Gaussian factor, $\exp\left({-\sum_k \frac{|z_k|^2}{4l_B^2}}\right)$, is sometimes not explicitly written since it is always the same factor regardless of the specific form of the wave function.

(ii) One of the most important lessons obtained from the study of liquid Helium is that the strongly correlated many-body wave function can be well described by the product of two-body wave functions so long as the two-body correlation is taken care of as accurate as possible. 
It turns out that higher-body correlations can be ignored to a good approximation.
Applying this idea to the FQHS problem, one can write the following many-body wave function for the FQHSs:
\begin{align}
\Psi = \prod_{i<j} f(z_i-z_j) e^{-\sum_k \frac{|z_k|^2}{4l_B^2}},
\end{align}
where $f(z)$ must be a polynomial of $z$ to satisfy the holomorphicity condition discussed above.
Note that we have dropped the argument of the many-body wave function, i.e., $z_1,z_2,\cdots,z_N$, for simplicity.
Incidentally, the function $f$ is generally called the Jastrow factor named after Jastrow, who considered this type of the wave function for the first time.

(iii) Since electrons are fermions,  $f(z)$ should be an odd function with respect to the sign change of $z$: $f(-z)=-f(z)$.

(iv) The simplest Jastrow factor incorporating all the above ideas is the power function of $z$ with an odd power: $f(z)=z^{2p+1}$ with $p$ being an integer. 
That is, the resulting many-body wave function is given as
\begin{align}
\Psi = \prod_{i<j} (z_i-z_j)^{2p+1} e^{-\sum_k \frac{|z_k|^2}{4l_B^2}},
\end{align}
which is none other than the Laughlin wave function.

(v) By inspecting the form of the Laughlin wave function, one can find that the size of the electron liquid described by the Laughlin wave function is set by the maximum power of any particular electron coordinate $z_i$, which is of course the same for all electrons.
Specifically, the maximum power is $m_{\rm max}=(2p+1)(N-1)$, which also defines the total number of the available orbitals for electrons participating in the Laughlin state.
Meanwhile, the filling factor $\nu$ is the ratio between the total number of electrons and available orbitals:
\begin{align}
\nu=\frac{N}{m_{\rm max}}=\frac{N}{(2p+1)(N-1)} \rightarrow \frac{1}{2p+1} ,
\end{align}
where the last expression is obtained in the thermodynamic limit of $N \rightarrow \infty$. 
In summary, the Laughlin wave function is defined at filling factor $\nu=1/(2p+1)$.

\begin{figure*}[t]
\begin{center}
\includegraphics[width=1.0\columnwidth,angle=0]{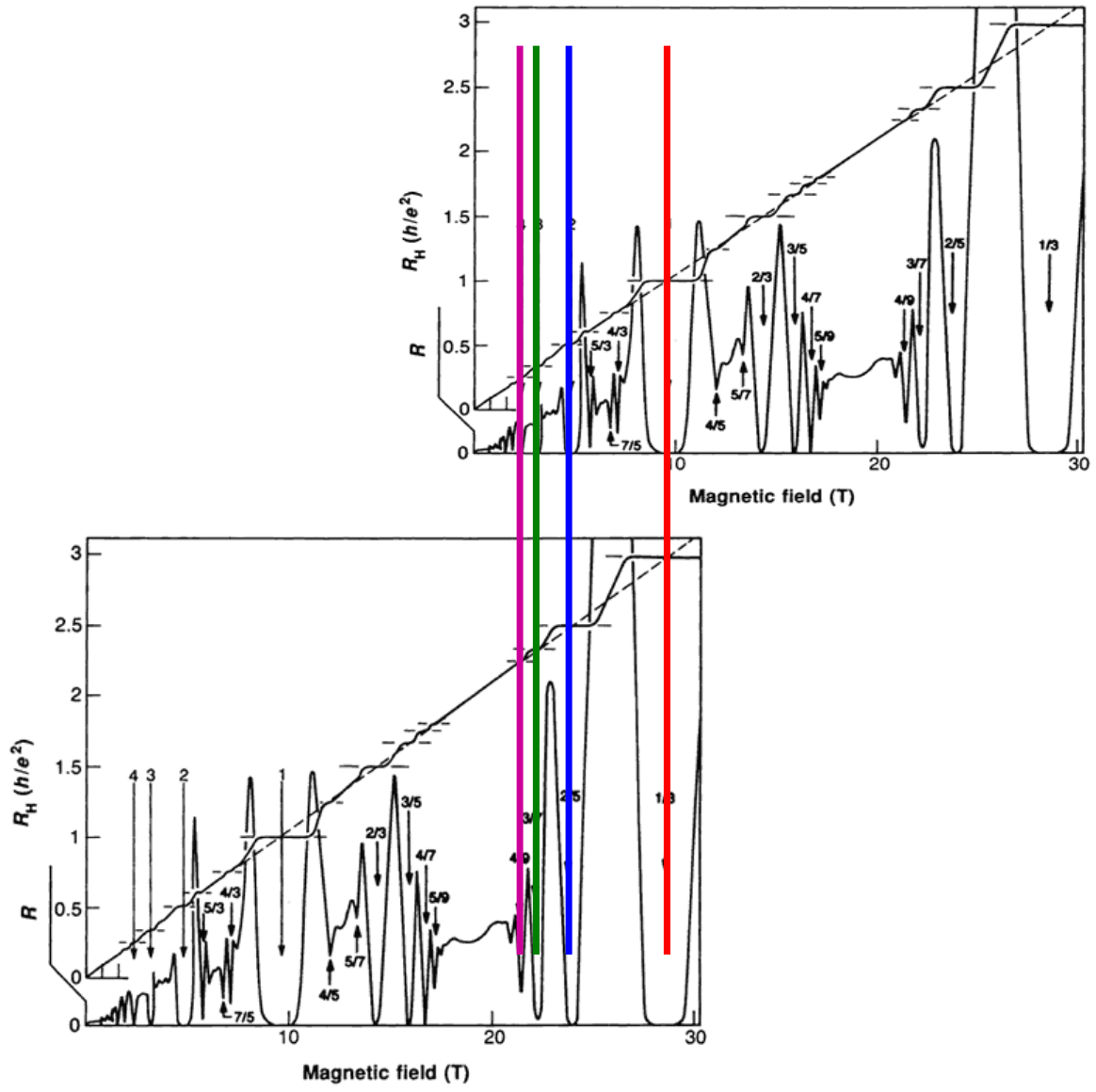}
\caption{
Mapping between the FQHSs of electrons at $\nu=n/(2n+1)$ and the IQHSs of CFs at $\nu^*=n$.
It is important to note that the top and bottom experimental plots are exactly identical except that the latter is translated to the left-hand side indicating that the magnetic field is subtracted by the constant amount equal to $2\rho\phi_0$ with $\rho$ being the electron density and $\phi_0$ being the flux quantum.
The experimental plot is taken from the press release of the Nobel Prize in Physics in 1998.
}
\label{fig:CF_mapping}  
\end{center}
\end{figure*}   

\begin{figure*}[t]
\begin{center}
\includegraphics[width=1.3\columnwidth,angle=0]{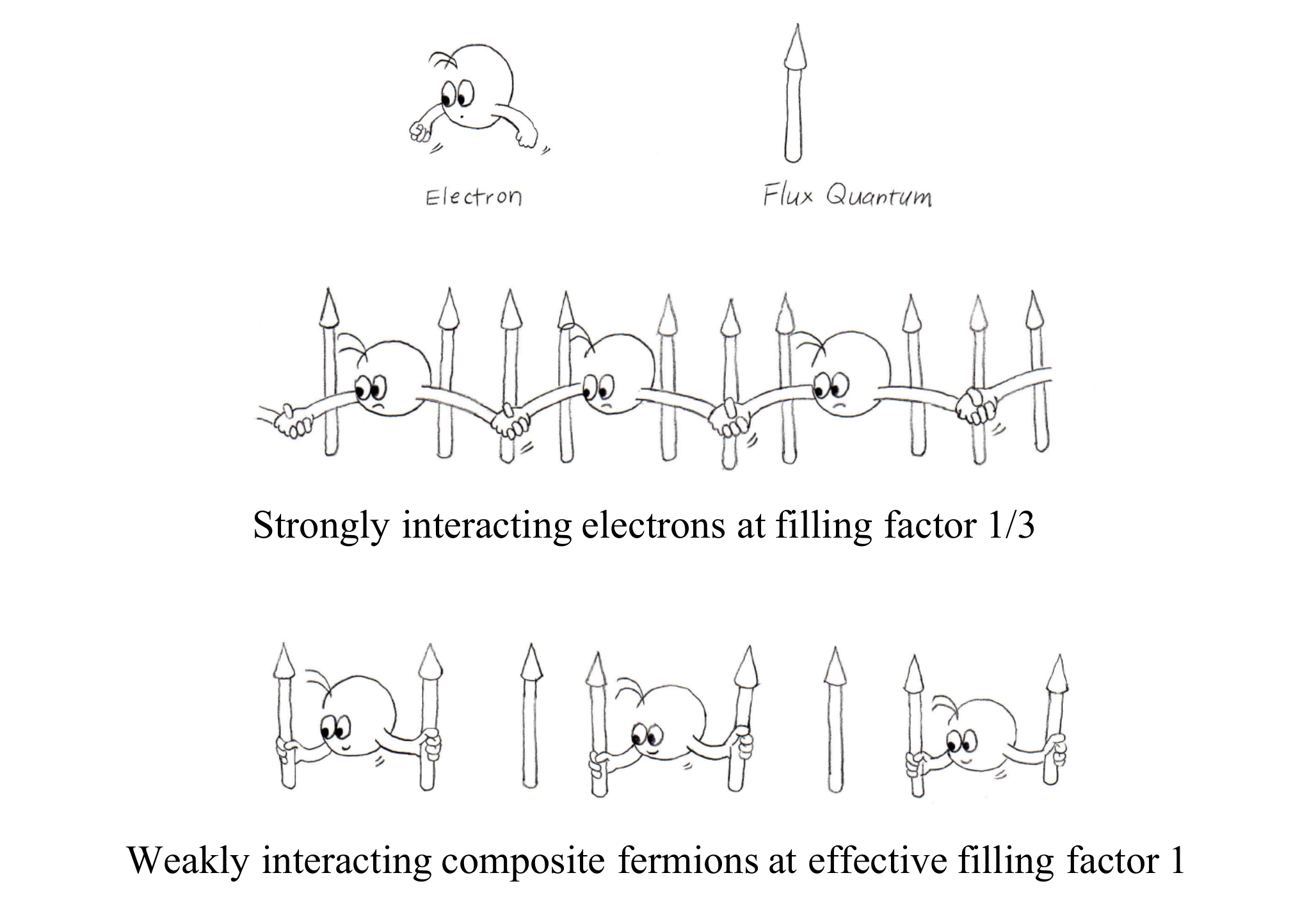}
\caption{
Cartoon picture of the CF theory.
This cartoon provides a humorous explanation for the formation of CFs.
Initially, electrons are strongly interacting with each other at filling factor $\nu=1/3$.
To reduce the Coulomb interaction energy, electrons are transformed into CFs by capturing, or grabbing two flux quanta nearby.
Consequently, CFs feel only the residual magnetic fields at effective filling factor $\nu^*=1$. 
Note that the (effective) filling factor is simply the ratio of the number of electrons (CFs) to that of flux quanta.
Incidentally, this cartoon was originally drawn by the current author of this review as an illustration inserted in the Ph.D. thesis of Rajiv Kamilla in 1997.
}
\label{fig:CF_cartoon}  
\end{center}
\end{figure*}   

It was soon discovered, however, that, in addition to the Laughlin sequence $\nu=1/(2p+1)$, various other FQHSs are obtained at the filling factors summarized by the following formula:
\begin{align}
\nu=\frac{n}{2pn\pm1} ,
\label{eq:Jain_sequence}
\end{align} 
where $n$ and $p$ are both integers.
Some FQHSs are related with others via the particle-hole symmetry such that $\nu=2-n/(2pn\pm1)$.

Eventually, the entire sequence of these FQHSs was explained by the composite fermion (CF) theory, which was put forward by Jain~\cite{Jain1989,Jain_Book}.
The CF theory provides a unification of the IQHSs and FQHSs via the key principle that there is a new quasiparticle called the CF, which is the bound state between an electron and an even number of vortices. 
Since a vortex can be roughly regarded as one magnetic flux quantum, this means that CFs experience only the residual magnetic field that is the difference between the external and the captured magnetic fields.
That is, the effective magnetic field experienced by the CF, $B^*$, is given as
\begin{align}
B^*= B-2p\rho \phi_0 ,
\label{eq:B_eff}
\end{align} 
where $2p$ denotes the number of vortices captured by a composite fermion, and $\rho$ is the electron density.

Defined as the ratio between the total number of electrons and flux quanta, the filling factor of electrons is related with that of CFs as follows:
\begin{align}
\nu= \frac{\rho}{B/\phi_0} = \frac{\rho}{B^*/\phi_0+2p\rho}=\frac{\nu^*}{2p\nu^*+1}.
\label{eq:nu_nu*}
\end{align}
If CFs fill an integer number of the effective Landau levels, i.e., forming an IQHS at $\nu^*=n$, then the filling factor for the FQHS of electrons is given as 
\begin{align}
\nu=\frac{n}{2pn+1} ,
\end{align}
which is a special case of Eq.~\eqref{eq:Jain_sequence} with the positive sign chosen in the denominator.
See Fig.~\ref{fig:CF_mapping} for a schematic diagram showing the mapping between the FQHSs of electrons at $\nu=n/(2n+1)$ (with $p=1$) and the IQHSs of composite fermions at $\nu^*=n$.
See also Fig.~\ref{fig:CF_cartoon} for a cartoon picture of the CF theory.

The negative sign is chosen if vortices are attached to CFs inversely.
The inverse vortex attachment is necessary when $B$ and $B^*$ have different signs.
In this situation, CFs fill $n$ effective Landau levels with the opposite residual magnetic field, i.e., $\nu^*=-n$, so that Eq.~\eqref{eq:nu_nu*} becomes
\begin{align}
\nu=\frac{-n}{-2pn+1}=\frac{n}{2pn-1} .
\end{align}
Incidentally, the filling factor sequence in Eq.~\eqref{eq:Jain_sequence} is called the Jain sequence.

Microscopically, the CF wave function can be written as follows:
\begin{align}
\Psi_\nu={\cal P}_{\rm LLL} \prod_{i<j} (z_i-z_j)^{2p} \Psi_{\nu^*} ,
\label{eq:CF_wave_function}
\end{align} 
where the Jastrow factor, $\prod_{i<j} (z_i-z_j)^{2p}$, play the role of attaching $2p$ vortices to CFs at the effective filling factor $\nu^*$.
Above, ${\cal P}_{\rm LLL}$ denotes the LLL projection operator, which is necessary since  $\Psi_{\nu^*}$ can in general contain $\bar{z}$ as well as $z$.
Note that, corresponding to $\nu^*=1$ and $p=1$, Eq.~\eqref{eq:CF_wave_function} reproduces the Laughlin wave function at $\nu=1/3$ since $\Psi_{\nu^*=1}=\prod_{i<j}(z_i-z_j)$, which is nothing but the Slater determinant for the fully filled LLL, also known as the Vandermonde determinant.
It has been shown that Eq.~\eqref{eq:CF_wave_function} provides very accurate wave functions for the exact Coulomb ground states, which are obtained via exact diagonalization of various finite-size systems~\cite{Jain_Book}.

Now, an important question is how to translate the CF wave function in Eq.~\eqref{eq:CF_wave_function}, which is written in the continuum, to a corresponding analog in the lattice.

{\bf Fractional Chern insulator.}
To find the lattice analog of the CF wave function, it is convenient first to consider the CF wave function in the Landau gauge, which is amenable for the application of the periodic boundary condition.
Since it is not straightforward to write the CF wave function in the Landau gauge at general filling factors~\cite{Hermanns2013}, here, we focus on the Laughlin wave function only.
The Laughlin wave function is written on a cylinder with the Landau gauge as follows~\cite{Rezayi1994}:
\begin{align}
\Psi_{\rm Laughlin}= \prod_{i<j} 
\left( 
e^{\frac{2\pi}{L_y}z_i} -e^{\frac{2\pi}{L_y}z_j}
\right)^{2p+1}
e^{-\sum_k \frac{x_k^2}{2l_B^2}} ,
\label{eq:Laughlin_on_cylinder}
\end{align}
where it is used that the cylinder is finite along the $y$ direction with the length being $L_y$, which means that the momentum in the $y$ direction is quantized in units of $2\pi/L_y$.

It is important to note that Eq.~\eqref{eq:Laughlin_on_cylinder} can be obtained by using a similar logic used to derive the Laughlin wave function in the circular gauge in Eq.~\eqref{eq:Laughlin}. 
That is, first, the LLL eigenstates can be written as follows:
\begin{align}
\psi_m({\bf r}) \propto e^{\frac{2\pi m}{L_y}z} e^{-\frac{x^2}{2l_B^2}} , 
\label{eq:LLL_eigenstates_on_cylinder}
\end{align}
which can be obtained from Eq.~\eqref{eq:LL_eigenstates} by setting $n=0$ and $k_y=2\pi m/L_y$.
Then, notice that $e^{2\pi m z/L_y}$ can be rewritten as  $g^m(z)$ with $g(z)=e^{2\pi z/L_y}$.
If so, the same logic used to derive Eq.~\eqref{eq:Laughlin} can generate the following Jastrow factor type of the wave function:
\begin{align}
\Psi= \prod_{i<j} \left[g(z_i)-g(z_j)\right]^{2p+1} e^{-\frac{x^2}{2l_B^2}} ,
\end{align}
which is nothing but the form given in Eq.~\eqref{eq:Laughlin_on_cylinder}.
Here, it is important to note that the above form is possible since the LLL wave function can be written as a polynomial of $g(z_i)$ similar to the situation in the circular gauge, where it can be written as a polynomial of $z_i$.

Now, it is possible to translate the Laughlin wave function by using a basis mapping from the LLL wave functions to the so-called hybrid Wannier functions~\cite{Qi2011}.
The hybrid Wannier function is a localized wave packet along one direction, while a plane wave in the other.
Referring detailed discussions for a concrete mathematical form of the hybrid Wannier function to Ref.~\cite{Qi2011}, let us simply denote the hybrid Wannier function as $\psi^{\rm hy\mbox{-}Wan}_m$ with $m$ denoting the momentum along the $y$ direction.
An important point here is that the hybrid Wannier function $\psi^{\rm hy\mbox{-}Wan}_m$ can be one-to-one mapped to the LLL wave function $\psi_m$ in Eq.~\eqref{eq:LLL_eigenstates_on_cylinder}.

Specifically, it is in principle possible to expand the Laughlin wave function in Eq.~\eqref{eq:Laughlin_on_cylinder} as follows:
\begin{align}
\Psi_{\rm Laughlin}= \sum_{\{m_i\}} {\cal C}_{\{m_i\}} {\cal A} \left[ \psi_{m_1}({\bf r}_1)\cdots\psi_{m_N}({\bf r}_N) \right].
\end{align}
Then, the FCI wave function for the Laughlin state can be written by just replacing $\psi_m({\bf r})$ by $\psi^{\rm hy\mbox{-}Wan}_m(\bf r)$:
\begin{align}
\Psi^{\rm FCI}_{\rm Laughlin}= \sum_{\{m_i\}} {\cal C}_{\{m_i\}} {\cal A} \left[ \psi^{\rm hy\mbox{-}Wan}_{m_1}({\bf r}_1)\cdots\psi^{\rm hy\mbox{-}Wan}_{m_N}({\bf r}_N) \right],
\label{eq:FCI_Laughlin}
\end{align}
where it is important to note that the same amplitudes ${\cal C}_{\{m_i\}}$ are used for both wave functions.
It was found that the wave function in Eq.~\eqref{eq:FCI_Laughlin} has a reasonably high overlap with the exact ground state of the model Hamiltonian defined in a nearly flat Chern band with next-nearest neighbor repulsive interaction~\cite{Wu2012}.

Despite reasonable agreements with exact diagonalization results, however, there are some serious issues in the above approach. 
First, the gauge is not uniquely defined in the basis mapping process.
As a consequence, one need a certain gauge fixing procedure, which is unfortunately somewhat arbitrary. 
Second, the specific form of the hybrid Wannier function is also chosen somewhat arbitrarily.
As a matter of principle, any function, which is localized in one direction and plane-wave-like in the other, would be sufficient.

The second issue is actually related with the fundamental difference between the Landau level and the Chern band.
While the Landau level can be in some sense regarded as a flat Chern band, there is also an important difference between the two. 
The Landau level eigenstates have a natural length scale called the magnetic length, which is determined by the strength of the magnetic field.
Meanwhile, the energy eigenstates in the Chern band are fundamentally the plane waves, or the Bloch states with both momenta in the $x$ and $y$ directions being good quantum numbers.
Therefore, there is no length scale.
It is important to note that the hybrid Wannier function is not the energy eigenstate. 
Of course, the interaction can introduce a new length scale via spontaneous symmetry breaking.
Unfortunately, if so, the ground state would be some kind of the charge density wave (CDW) state rather than the quantum Hall state.

In summary, while it is likely that the Laughlin-like state can be obtained in the fractionally filled Chern band, there are still some issues to be resolved in order for the FCI to exist both conceptually and experimentally.

{\bf Beyond the independent bipartite fractional topological insulator.}
The 2D topological insulator is composed of two independent copies of the Chern insulator with opposite Chern numbers for different spin species, preserving the time-reversal symmetry as a whole. 
Naturally, the FTI has been proposed as being composed of two independent copies of the FCI~\cite{Bernevig2006_PRL,Levin2009,Maciejko2010,Santos2011,Levin2011,Lu2012,Chen2012,Levin2012,Klinovaja2014,Repellin2014,Furukawa2014,Stern2016}.
Let us call this type of the FTI the independent bipartite FTI.

A question is if this scenario is generically true in the physically realistic situation, where the electron-electron interaction has the same strength regardless of spin.
It is important to note that the FCI of each spin species is induced by the electron-electron interaction.
Therefore, the independent bipartite FTI can be in principle obtained in an artificial limit, where the interspin interaction is much weaker than the intraspin interaction. 
It has been indeed shown that the independent bipartite FTI can be stable up to a certain strength of the interspin interaction relative to the intraspin interaction~\cite{Chen2012}.
Unfortunately, the independent bipartite FTI breaks down for the realistic interaction, where the interspin interaction has the same strength as the intraspin interaction.

In this context, it is interesting to investigate the true nature of the Coulomb ground state in the fractionally filled Landau levels with spin-dependent holomorphicity, i.e., electrons of one spin species reside in the holomorphic Landau level, while those of the other reside in the antiholomorphic counterpart.
It has been shown in a recent paper of the current author~\cite{Mukherjee2017} that the ground state is generally compressible and disordered except at half filling, where the filling factor of each spin species is a half. 
Surprisingly, an incompressible state at half filling is susceptible to an inherent spontaneous symmetry breaking, eventually leading to the spatial separation of different spins.
This means that, in general, the FTI cannot be described as an independent bipartite form.

{\bf Discussion}

I would like to conclude this review by discussing some of the future directions in the research of topological matter.

First, strongly correlated topological matter is expected to become more and more important.
Remembering that the FCI is a lattice analog of the FQHS, which is one of the most intriguing strongly correlated systems, the FCI would become an active research field if certain conditions for its experimental observation are met.
One of the most important experimental conditions is of course the existence of the (nearly) flat Chern band. 
Next, the Coulomb interaction must be sufficiently strong so that it can overcome the effects of band dispersion.
While there is a long way to go, it would be exciting to observe a fractionally quantized Hall resistance without applying an external magnetic field.

Second, while various types of topological matter have been found in real materials, an interesting direction to pursue is the artificial generation of topological matter, especially by applying a time-periodic operation to the system.
This operation has been dubbed as the Floquet engineering~\cite{Oka2018}.
One of the most notable examples in the Floquet engineering is the proposal for the generation of a Floquet topological insulator by irradiating graphene~\cite{Oka2009, Kitagawa2011, Kundu2014, Dehghani2015, Sentef2015, Mikami2016} with a circularly polarized light at high frequency.
Such a Floquet topological insulator is particularly interesting since it can provide an exact realization of the Haldane model~\cite{Haldane1988} or the Kane-Mele model~\cite{Kane2005} for a single spin species with the possibility of manipulating the Chern number via tuning the radiation electric field strength.
Also, it has been recently shown by the current author~\cite{Kim2018} that a Floquet topological semimetal with nodal helix can be generated by irradiating graphene with a circularly polarized light at {\it low} frequency.

{\bf Acknowledgement}

The author is grateful to Sutirtha Mukherjee for careful reading of the manuscript and providing various useful comments.



\end{document}